\documentclass[a4paper,twocolumn,11pt,accepted=2026-10-20]{quantumarticle}
\pdfoutput=1
\usepackage[utf8]{inputenc}
\usepackage[english]{babel}
\usepackage[T1]{fontenc}
\usepackage{amsmath}
\usepackage{hyperref}
\usepackage[nocompress]{cite}
\usepackage{tikz}
\usepackage{lipsum}
\usepackage{ragged2e}
\usepackage{braket}
\usepackage{dsfont}

\begin{document}

\title{Optimal phase estimation in the presence of correlated dephasing}

\author{Srijon Ghosh$^*$}
\affiliation{Faculty of Physics, University of Warsaw, Pasteura 5, 02-093 Warszawa, Poland}
\affiliation{Dipartimento di Ingegneria, Università di Palermo, Viale delle Scienze, 90128 Palermo, Italy}
\affiliation{Institute of Fundamental and Frontier Sciences, University of Electronic Sciences and Technology of China, Chengdu 611731, China}
% \thanks{These two authors contributed equally to this work.}
\author{Arkadiusz Kobus$^*$}
\affiliation{Faculty of Physics, University of Warsaw, Pasteura 5, 02-093 Warszawa, Poland}
\affiliation{Institute of Physics PAS, Aleja Lotników 32/46, 02-668 Warszawa, Poland}
% \thanks{These two authors contributed equally to this work.}

\author{Stanisław Kurdziałek}
\affiliation{Faculty of Physics, University of Warsaw, Pasteura 5, 02-093 Warszawa, Poland}

\author{Rafał Demkowicz-Dobrzański}
\affiliation{Faculty of Physics, University of Warsaw, Pasteura 5, 02-093 Warszawa, Poland}

\def\thefootnote{*}\footnotetext{These authors contributed equally to this work}

% \email{latex@quantum-journal.org}
% \homepage{http://quantum-journal.org}
% \orcid{0000-0003-0290-4698}
% \thanks{You can use the \texttt{\textbackslash{}email}, \texttt{\textbackslash{}homepage}, and \texttt{\textbackslash{}thanks} commands to add additional information for the preceding \texttt{\textbackslash{}author}. If applicable, this can also be used to indicate that a work has previously been published in conference proceedings.}
% \affiliation{Covestro Deutschland AG, Kaiser-Wilhelm-Allee 60, 51373 Leverkusen, Germany}
% \author{Marcus Huber}
% \affiliation{Institute for Quantum Optics \& Quantum Information (IQOQI), Austrian Academy of Sciences, Boltzmanngasse 3, Vienna A-1090, Austria}
% \orcid{0000-0003-1985-4623}
% \author{Cassandra Granade}
% \affiliation{Microsoft Research, Quantum Architectures and Computation Group, Redmond, WA 98052, USA}
% \author{Johannes Jakob Meyer}
% \affiliation{Dahlem Center for Complex Quantum Systems, Freie Universität Berlin, 14195 Berlin, Germany}
% \orcid{0000-0003-1533-8015}
% \author{Victor V. Albert}
% \affiliation{Institute for Quantum Information and Matter \& Walter Burke Institute for Theoretical Physics, Caltech, Pasadena, CA 91125, USA}
% \orcid{0000-0002-0335-9508}
\maketitle

\begin{abstract}
 We investigate optimal metrological protocols for phase estimation in the presence of correlated dephasing noise, including spin-squeezed state sensing strategies as well as parallel and adaptive protocols optimized using tensor-network-based numerical methods. The results are benchmarked against fundamental bounds obtained either via the latest quantum comb extension method or an optimized classical simulation method. We find that the spin-squeezed states offer practically optimal performance in the regime where phase fluctuations are positively correlated but can be outperformed by tensor-network-optimized strategies for negatively correlated fluctuations.
\end{abstract}

\section{Introduction}
\label{sec:intro}

A fundamental challenge in theoretical quantum metrology \cite{Giovannetti2004,Giovannetti2011,Demkowicz-Dobrzanski2015a,Degen2016,Pezze2018,Pirandola2018,Polino2020,Liu2022c,Jiao2023a,Liu2024s,Montenegro2024} lies in identifying the optimal estimation scheme that achieves the highest precision, when the sensing probes experience realistic noise. %In such practical settings, developing rigorous methods to establish the fundamental precision bounds of quantum sensing protocols and designing strategies capable of saturating these bounds remains a highly non-trivial task. 
Addressing this challenge is of profound importance, both from a fundamental as well as a practical perspective, as it impacts the advancement of practical quantum sensing technologies \cite{Pirandola2018, caves1981, rafal2015, Tsang2016, Alessandro2020, taylor2016}. 

%It can be shown \sk{that classically} the maximum attainable precision for the parameter of interest scales as $N^{-1/2}$, where $N$ is the number of resources used \cite{Giovannetti2011}. This is well-known as the standard quantum limit. 
While in ideal noiseless scenarios, proper use of quantum mechanical resources, such as coherence or entanglement, may lead to the famous Heisenberg scaling of precision \cite{ Giovannetti2004, Giovannetti2011, Ou1997, Dur2014, Zhou2017}, in the presence of 
noise, those quantum-enhanced gains, while still possible, are less spectacular, and require more careful design of the metrological protocols \cite{Zhou2017, Escher2011, Demkowicz-Dobrzanski2012, Demkowicz-Dobrzanski2014, Sekatski2016, Demkowicz-Dobrzanski2017, Zhou2020}
%However, an experimentally relevant situation is always affected by noise, and in such cases, the precise estimation of a parameter requires significant effort. 

Mathematically, the problem of identifying an optimal sensing strategy may be framed as a quantum channel estimation problem, where the parameter $\theta$ is encoded in a quantum channel $\Lambda_{\theta}$.
The problem is at the same time most challenging and interesting from a practical perspective, when the task involves the use of multiple quantum channels, which may be uncorrelated, correlated spatially, temporally, or both \cite{Matsuzaki2011,Chin2012,Macieszczak2015,Smirne2015a,Haase2017,Szankowski2014,Smirne2018,Tamascelli2020,Altherr2021,Yang2024j,Jeske2014,Layden2018,Czajkowski2019,Layden2019,Layden2020,Planella2022,Beaudoin2018,Riberi2022}. For uncorrelated channel estimation models, the problem may be formally regarded as solved, as both  asymptotically tight fundamental bounds are known as well as the optimal schemes saturating them  \cite{Fujiwara2008,Escher2011,Demkowicz-Dobrzanski2012, Koodynski201,Demkowicz-Dobrzanski2014,Sekatski2016, Zhou2017, Demkowicz-Dobrzanski2017,Zhou2019e,Zhou2020,Kurdzialek2023a}. 

In a more general scenario, the noise channels are likely to exhibit correlations. However, characterizing and analyzing correlated quantum channels is a non-trivial task, and the aforementioned bounds and strategies do not, in general, prove useful in this setting. However, more recently, a lot of progress has been made in developing methods to deal with correlated metrological noise models as well \cite{Yang2019, Altherr2021, Liu2023b, Liu2024, Kurdzialek2024, kurdzialek2025, das2025}. Still, universal constructions of optimal protocols are not known, while the efficiently computable bounds are not guaranteed to be tight in general. To the best of our knowledge, the most recent works \cite{Kurdzialek2024, kurdzialek2025} have focused on the toy model of binary correlated dephasing. Precisely, the two adjacent quantum channels are correlated via only two possible values of the rotation angles.

In this work, we move beyond the binary picture, allowing the correlations between quantum channels to vary continuously.
Precisely, we focus on a paradigmatic metrological model with correlated noise---phase estimation in the presence of correlated dephasing---where the character of noise may be varied continuously, changing it from anti-correlated noise, via uncorrelated (independent dephasing), up to the fully correlated one (collective dephasing). In this scenario, to derive the fundamental upper bounds of the Fisher information, we use \textit{classical simulation} (CS) \cite{Demkowicz-Dobrzanski2012, Koodynski201} method. At first, we calculate a tighter upper bound of the Fisher information by using CS method for the \textit{continuous correlated model} as well as by using the quantum comb extension method \cite{kurdzialek2025}. We apply the latest tensor-network-based numerical methods \cite{Kurdzialek2024, Liu2024}
to find the optimal parallel as well as adaptive protocols with up to $N \approx 30$ channels, and compare their performance with basic spin-squeezed state-based strategies, for different character of noise correlations by invoking the Rouwenhorst method of discretizing a continuous noise-model. 
% We derive a new upper bound for the model making the most out of the classical simulation method  as well as employ the recently developed method to compute the bounds using quantum comb extension method \cite{kurdzialek2025}.}
% Moreover, for a tighter upper bound, we use the channel extension method in the correlated scenario \cite{kurdzialek2025}. By using the Rouwenhorst method \cite{Kopecky2010}, a discretization prescription for an autro-regressive (AR) process, we approximate the continuous noise model as a probability distribution within a finite state space. This tool rigorously helps us to compute the Fisher information efficiently for a large number of channel uses. Nonetheless, we establish a lower bound and identify optimal estimation strategies, which are examined through an informed selection of probe states—such as spin-squeezed states—and through tensor-network-based optimization using matrix product states (MPS) and matrix product operators (MPO) for the initial probe state and the measurement scheme, respectively. Moreover, we introduce a multi-observable estimator for the spin squeeze state and analyze the effect of measurement optimization in the presence of correlated noise. Next, we will briefly discuss the aforementioned physical noise model that affects the spins of the probe state as a continuous random fluctuation in the sensing of a magnetic field.\\

The paper is organized as follows. In Sec. \ref{sec:physical model}, we formulate the model of phase estimation in the presence of correlated phase fluctuations. In Sec. \ref{sec:phase estimation} we analyze the performance of a parallel channel estimation strategy involving spin-squeezed states and simple measurements in the presence of correlated dephasing. In order to obtain a benchmark for optimality, 
we derive general bounds on precision with correlated noise using classical simulation method in Sec. \ref{sec:CS bounds}. In Sec. \ref{sec:tensor network}, we present numerical results on protocols optimized with the help of tensor network methods, which are supplemented by numerically obtained fundamental bounds 
obtained using the quantum comb extension method.
% In order to further investigate the potential of  squeezed-states based strategies, in Sec. \ref{sec:higherorder} we investigate the improvement of performance of the protocol thanks to measurements of higher-order spin-observables. 
Finally, conclusions are drawn in Sec. \ref{sec:conclusion}.

\section{Phase estimation under the correlated dephasing problem}
\label{sec:physical model}

\begin{figure}[t]
  %  \centering
    \includegraphics[width=\linewidth]{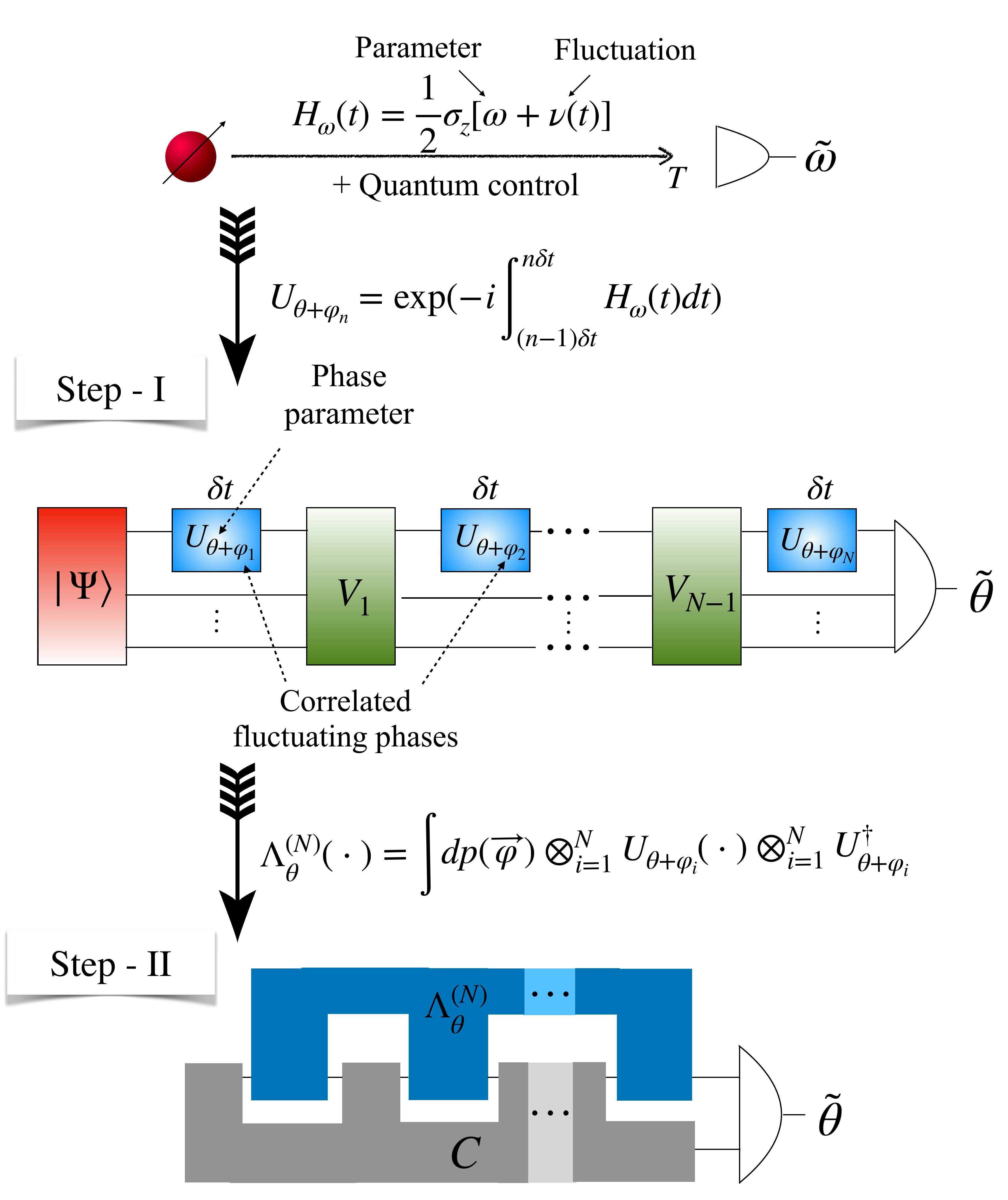}
    % \caption{ Schematic diagram of a metrological channel estimation protocol with continuous temporal correlations. The continuous-time frequency estimation problem is shown at the top, where $T$ denotes the time. \textbf{Step I} indicates the mapping of the aforementioned problem to a general \textit{discrete-time phase sensing problem} involving $N$ steps (middle), where each step represents sensing for a time $\delta t$ followed by a quantum control operation $V_{n}$, with $n= 1,2,\ldots, N-1$. \textbf{Step - II} indicates that both the probe sensing dynamics $\Lambda_\theta^{(N)}$ as well as the entire control protocol $C$ can be represented mathematically as quantum combs (bottom).}
    \caption{Schematic diagram of a metrological channel estimation protocol with continuous temporal correlations. The continuous-time frequency estimation problem is illustrated at the top, where $T$ denotes the total evolution time. \textbf{Step I} represents the mapping of this problem onto a general \textit{discrete-time phase sensing protocol} comprising $N$ steps (middle), where each step involves sensing for a duration $\delta t$, followed by a quantum control operation $V_{n}$, with $n = 1, 2, \ldots, N-1$. \textbf{Step II} involves averaging over noise processes and represents mathematically both the resulting sensing dynamics $\Lambda_\theta^{(N)}$ and the overall control protocol $C$ as quantum combs (bottom).}
    \label{fig:schematic}
\end{figure}

Let us start with a simple physical motivation for the model, which will be the main focus of this paper. 
Imagine sensing of a time-varying magnetic field $B(t)$
that is fluctuating around some stable mean value $B_0$. Consider an elementary two-level quantum system that couples to the magnetic field via a standard  Hamiltonian:
\begin{equation}
H_{\omega}(t) = \frac{1}{2}\mu B(t) \sigma_z  = \frac{1}{2}\omega(t) \sigma_z,
\end{equation} where 
$\mu$ is the coupling constant (magnetic moment), while $\omega(t)$ is the corresponding Larmor frequency. We 
may therefore view the problem as effectively a frequency estimation problem,
\begin{equation}
    \omega(t) = \omega + \nu(t),
\end{equation}
where the goal is to estimate $\omega$ in the presence of frequency fluctuations denoted here by $\nu(t)$. We allow the system to be subject to arbitrary controls during sensing and, in principle, also entangled with ancillary systems; see Fig.~\ref{fig:schematic}.

%[width= 16cm, height = 11cm] 
In order to effectively analyze the model, we assume sensing steps last $\delta t$ and are intertwined with control operations $V_n$. 
We may then view the problem as a phase estimation problem, with sequential application of unitary operations,
\begin{equation}
U_{\theta + \varphi_n} = \textrm{e}^{-\text{i}\frac{1}{2} (\theta + \varphi_n)\sigma_z }, \ \theta = \omega \delta t, \  \varphi_n =\int_{t_n}^{t_{n+1}}\nu(t)\text{d}t,
\end{equation}
 where $\varphi_n$ represents the total phase fluctuations accumulated from time $t_n=(n-1)\delta t$ to $t_{n+1} = n \delta t $. The goal is to optimally estimate phase $\theta$, which directly translates to estimation of $\omega$. What makes the problem non-trivial is that we allow for arbitrary controls that may be applied between the applications of the unitaries.
 If the fluctuating phases $\varphi_n$ are independent, we would end up with the standard uncorrelated dephasing model, for which it is known that a squeezed-state-based sensing protocol performs optimally  \cite{Huelga1997, Orgikh2001, Escher2011, Demkowicz-Dobrzanski2012}.  The problem becomes highly non-trivial, however, if 
 fluctuating phases $\varphi_n$ manifest correlations.
Better understanding of this case is the main purpose of this paper.

We will consider the simplest model that is sufficient to study the effect of dephasing correlations,  where the vector of subsequent phases $\vec \varphi = [\varphi_1, \varphi_2,...]^T$ is normally distributed,  
\begin{align}
\label{eq:gaussian}
    \vec{\varphi}&\sim\mathcal{N}(0,\Sigma),\quad\Sigma_{ij}=\sigma^2c^{|i-j|},
\end{align}
with $\Sigma$ being the covariance matrix,  $c\in(-1,1)$ being the noise correlation parameter and  $\sigma$ being the noise strength. 
This assumption makes the model Markovian, 
% \sk{[maybe change 'as' to 'and.] The fact that we can write a formula for a conditional probability doesn't prove markovianity]} \sg{Yes, and in that case we should rewrite the sentence completely.} 
and the conditional probability of observing phase $\varphi_{n+1}$ given $\varphi_{n}$ in the previous step reads:
\begin{equation}
\varphi_{n+1}|\varphi_n \sim   c \varphi_n+\mathcal{N}\left(0,{\sigma^2}{(1-c^2)}\right),
\end{equation}
A similar type of model would arise 
when integrating a natural frequency estimation model, where the frequency fluctuations are governed by the Ornstein-Uhlenbeck process \cite{OrnsteinUhlenbeck1930,stockton2004robust,petersen2006estimation,amoros2021noisy,amoros2025tracking}, see Appendix~\ref{sec:ou}---in this case though, one would be restricted to positive correlations only, $c \geq 0$, while here we allow for negative correlations as well---such correlation may appear in underdamped models with overshoot-type relaxations or in other oscillating models \cite{stelzer2011carma,grahovac2026superpositions}.

We may now state the problem formally, as a channel estimation problem, where the  $\theta$ parameter to be estimated is encoded in the effective channel:
\begin{equation}
\label{eq:N-partite channel}
    \Lambda^{(N)}_\theta( \cdot )=\int\text{d}p(\vec{\varphi})\left(\bigotimes_{n=1}^N U_{\theta+\varphi_n}\right) (\cdot )\left(\bigotimes_{n=1}^NU_{\theta+\varphi_n}\right)^\dagger,
\end{equation}
where $p(\vec{\varphi})$ is the $N$-variate Gaussian distribution \eqref{eq:gaussian}.
This channel may be viewed as acting on an $N$-qubit input probe state $\rho^{(N)}$, in which case it will represent a parallel sensing strategy, where the output state is $\rho_\theta^{(N)} = \Lambda^{(N)}_\theta(\rho^{(N)})$. More generally, it may be viewed as a quantum comb \cite{Chiribella2009}, with $N$-qubit inputs and $N$-qubit outputs, that may be used in a general adaptive strategy as depicted in Fig.~\ref{fig:schematic}. In the latter case, the probe-control strategy will be mathematically described by a complementary quantum comb $C$, while the output state of the protocol will be obtained via a concatenation of the two combs (the link product) $\rho_\theta^{(N)} = \Lambda_\theta^{(N)} \star C$ \cite{Chiribella2009, Kurdzialek2024}.

In this work, we will quantify the performance of the protocol via the \emph{quantum Fisher information (QFI)} \cite{helstrom1976quantum, braunstein1994statistical} of the protocol output state $\mathcal{F}_Q(\rho_\theta^{(N)})$. The QFI is defined as:
\begin{equation}
    \mathcal{F}_{Q}(\rho_\theta) = \mathrm{Tr} (\rho_{\theta} L_{\theta}^2),\quad 2\dot{\rho}_\theta=L_\theta\rho_\theta+\rho_\theta L_\theta,
\end{equation}
where $L_{\theta}$ denotes the symmetric logarithmic derivative (SLD)  and $\dot{X}=\partial_\theta X$. 
QFI provides the ultimate lower bound on the estimation variance of $\theta$ through the quantum Cramér-Rao (CR) inequality:
\begin{equation}
    \mathrm{Var}(\hat\theta) \ge  \frac{1}{\mathcal{F}_{Q}(\rho_{\theta})},
\end{equation}
valid for any measurements and locally unbiased estimators.

In the case of parallel protocols, the optimal 
protocol will correspond to the choice of the $N$-probe input state (potentially also entangled with ancillary systems) for which 
the output QFI is maximized
\begin{equation}
\label{eq:qfiparallel}
    \mathcal{F}_{Q,\textrm{parallel}}^{(N)}= \max_{\rho^{(N)}} \mathcal{F}_{Q}[\Lambda_{\theta}^{(N)} (\rho^{(N)})].
\end{equation}
More generally, in the case of adaptive strategies, the maximization should be performed over all legitimate quantum combs $C$:
\begin{equation}
\label{eq:qfiadaptive}
    \mathcal{F}_{Q,\textrm{adaptive}}^{(N)}= \max_{C} \mathcal{F}_{Q}[\Lambda_{\theta}^{(N)} \star C].
\end{equation}
If ancillary systems involved in the adaptive strategy are not restricted, the parallel strategy may always be regarded as a special case of the adaptive strategy, and so $\mathcal{F}_{Q,\textrm{adaptive}}^{(N)} \geq \mathcal{F}_{Q,\textrm{parallel}}^{(N)}$.

Performing exact optimizations in (\ref{eq:qfiparallel},\ref{eq:qfiadaptive}), quickly becomes infeasible with growing $N$, due to the exponential growth of the total Hilbert space dimension involved.
In this case, one needs to consider either some particular educated guessed protocols, or perform numerical optimization using efficient descriptions involving tensor-network methods, modeling entangled states as matrix product states (MPS) in parallel protocols \cite{Chabuda2020}, or decomposing general quantum combs into elementary steps, connected via ancillary systems of small dimensions \cite{Kurdzialek2024, Liu2024}.  
In the end, these results need to be accompanied by upper bounds, that allow to assess how big the potential gap is between the protocols found and the best possible ones. In the next section, we analyze the performance of the basic parallel phase estimation strategy, where quantum precision enhancement is achieved thanks to spin-squeezing, which is known to be optimal in for the problem of phase estimation in the presence of uncorrelated dephasing \cite{Huelga1997, Escher2011, Demkowicz-Dobrzanski2012, PhysRevA.47.5138,MA201189,kobus2025asymptoticallyoptimaljointphase}.

\section{Phase estimation using squeezed states}
\label{sec:phase estimation}
Consider the $ N$- qubit input state in the parallel protocol to be the one-axis twisting \emph{spin-squeezed (SS) state} \cite{MA201189}, $\rho^{(N)} = \ket{\textrm{SS}}\!\bra{\textrm{SS}}$:
\begin{equation}
\label{eq:ss}
    |\text{SS}\rangle=e^{\mathrm{i}\varepsilon J_x}e^{-\mathrm{i}\chi J_z^2}|+\rangle^{\otimes N},\quad J_\alpha=\frac{1}{2}\sum_{n=1}^N\sigma_\alpha^{(n)},
\end{equation}
where $\sigma_\alpha^{(n)}$ represents a Pauli matrix acting on the $n$-th qubit, and the squeezing strength $\chi= N^{-3/4}$ as well as the additional state rotation $\varepsilon=\pi/2+N^{-1/4}$ are chosen to decrease with $N$ in a way that guarantees optimal asymptotic performance in the standard Ramsey interferometry protocol in the presence of uncorrelated dephasing \cite{PhysRevA.47.5138,MA201189,kobus2025asymptoticallyoptimaljointphase}. 
%being the Pauli matrix acting on $n$-th qubit.
The important characteristic of this state, which in the end allows for the optimal phase estimation, is its first and second moments:
\begin{align}
\label{eq:OAT_a}
\langle\sigma_x^{(n)}\rangle&\simeq 1, & \langle\sigma_y^{(n)}\rangle &=0, & \langle \sigma_\alpha^{(n)}\sigma_\alpha^{(n)} \rangle & = 1 \\
\label{eq:OAT_b}
\langle\sigma_x^{(i)}\sigma_x^{(j)}\rangle&\simeq 1, & \langle\sigma_y^{(i)}\sigma_y^{(j)}\rangle & \simeq\frac{-1}{N}, & \langle\sigma_x^{(i)}\sigma_y^{(j)}\rangle& \simeq0,
\end{align}
for $i\neq j$. Throughout this paper we use the symbol $\simeq$ to mean that such approximation correctly predicts first-order asymptotic behavior of QFI or precision in the limit $N\to\infty$.

%Let us also compute the moments for output state $\Lambda_\theta^{(N)}(|OAT\rangle\langle OAT|)$:
%\sk{I understand that this is for uncorrelated noise? It is not clear from this notation. Also, we should skip this uncorrelated part and focus on correlated from the beginning. We should just keep the final result for variance for uncorrelated case.}
%\begin{align}
%\label{eq:OAT1}
%\langle\sigma_x^{(n)}\rangle\simeq\eta\cos\theta,\quad\langle\sigma_y^{(n)}\rangle\simeq\eta\sin\theta,\\
%\label{eq:OAT2}
%\langle\sigma_y^{(i)}\sigma_y^{(j)}\rangle\simeq\eta^2\left(\sin^2\theta-\frac{\cos^2\theta}{N}\right),\quad i\neq j.
%\end{align}
After the state is evolved in parallel through channel $\Lambda_\theta^{(N)}$, the value of $\theta$ is inferred based on the measurement of the $J_y$ observable on the output state $\rho_\theta^{(N)} = \Lambda_\theta^{(N)}(\rho^{(N)})$.
 We may assess the 
estimation variance of this procedure via the standard linear error propagation formula:
\begin{equation}
\label{eq:old_estimator}
    \hat{\theta}=\frac{J_y}{\partial_\theta\langle J_y\rangle},\quad \text{Var}(\hat{\theta})=\frac{\text{Var}(J_y)}{|\partial_\theta\langle J_y\rangle|^2},
\end{equation}
where $\hat{\theta}$ is the operator representing the local unbiased estimator, whose eigenvalues are the values of $\theta$ to be estimated.
%Using the standard Heisenberg picture methods \cite{MA201189}, we may evolve back 

%and by substituting OAT moments (\ref{eq:OAT1},\ref{eq:OAT2}) we obtain:
%\begin{align}
%    \partial_\theta\langle J_y\rangle&=\langle J_x\rangle\simeq \frac{N\eta}{2},\\
%    \text{Var}(J_y)&=\frac{1}{4}\sum_{i,j=1}^N\langle\sigma_y^{(i)}\sigma_y^{(j)}\rangle\simeq \frac{N(1-\eta^2)}{4},\\
 %   \text{Var}(\hat\theta)&\simeq\frac{1-\eta^2}{N\eta^2},
%\end{align}
%when $\theta=0$, so we saturate asymptotically the CR bound \eqref{eq:uncorrelated bound}.

Let us analyze the performance of the above protocol using the SS probe state and $J_y$ measurement for the correlated noise. 
We will use standard Heisenberg picture methods to compute the first and second moments of the $J_y$ observable by evolving the operators back through the channel and then compute the expectation values on the initial state \eqref{eq:ss} \cite{MA201189}.  Without loss of generality, we assume the estimation is performed around $\theta=0$.
In order to compute the estimation variance, we need to compute the following:
\begin{align}
    \partial_\theta\langle J_y\rangle&=\langle J_x\rangle = \frac{1}{2}\sum_{i=1}^N \langle \sigma_x^{(i)}\rangle, %\rangle\simeq \frac{N\eta}{2},
    \\
    \text{Var}(J_y)&=\frac{1}{4}\sum_{i,j=1}^N\langle\sigma_y^{(i)}\sigma_y^{(j)}\rangle.
\end{align}
For compactness of notation, in what follows all operator expectation values, $\langle \hat{A} \rangle$, are understood to be performed on the output state $\rho_\theta^{(N)}$, while when
we write $\langle f(\vec\varphi) \rangle$ with $f(\vec\varphi)$ being some function of classical random variable $\vec\varphi$, we understand that this represents averaging with respect to Gaussian probability distribution \eqref{eq:gaussian}.

With the help of SS properties \eqref{eq:ss} the 
relevant moments of Pauli matrices can be expressed in terms of
statistical properties of fluctuating phases 
\begin{align}
\label{eq:sigmx}
\langle\sigma_x^{(n)} \rangle & \simeq \langle \cos \varphi_n \rangle = e^{-\sigma^2/2} = \eta\\ \langle\sigma_y^{(i)}\sigma_y^{(j)}\rangle&\simeq\langle\sin\varphi_i\sin\varphi_j\rangle-\frac{1}{N}\langle\cos\varphi_i\cos\varphi_j\rangle,
\end{align}
for $i\neq j$, where
\begin{align}
    % \langle\cos\varphi_k\rangle=e^{-\sigma^2/2}\\
    \langle\sin\varphi_i\sin\varphi_j\rangle&=e^{-\sigma^2}\sinh\left(\sigma^2c^{|i-j|}\right),\\
    \langle\cos\varphi_i\cos\varphi_j\rangle&=e^{-\sigma^2}\cosh\left(\sigma^2c^{|i-j|}\right).
% ,\quad D=-\log\eta^2,
\end{align}
In \eqref{eq:sigmx} we have introduced $\eta = e^{-\sigma^2/2}$ that represents the effective shrinking factor of the Bloch vector of a single qubit state under the dephasing process. 
%and $\theta=0$. We now assume $\vec{\varphi}$ is homogeneous process satisfying the Markov property: \sk{ I think we should write a general formula \eqref{eq:lower} directly after (25), and then show results for two cases: AR process and Markov process. Maybe we don't have to write a general results for Markov process, but just for binary fluctuations to make the clear meaning of $c$ parameter. }
%\begin{equation}
 %   p(\vec{\varphi})=p(\varphi_1)p(\varphi_2|\varphi_1)p(\varphi_3|\varphi_2)\dots p(\varphi_N|\varphi_{N-1}).
%\end{equation}
%When the Markov chain is non-degenerate (there is non-zero probability of transfer between any two states in a finite time), the correlations decay exponentially: \sk{What is $c$? If we do not define it, it's too abstract.   }
%\begin{equation}
 %   \begin{matrix}\langle \sin\varphi_i\sin\varphi_j\rangle\approx A c^{|i-j|}\\\langle\cos\varphi_i\cos\varphi_j \rangle\approx\eta^2\qquad\phantom{.}\end{matrix}\text{ for some $A,c$ if $|i-j|\gg 1$,}
%\end{equation}
We may now obtain the formula for the estimation variance, which for simplicity we write in the form valid in the asymptotic $N \rightarrow \infty$ regime (assuming $|c|<1$ so that correlations die exponentially):
\begin{equation}
\label{eq:lower}
    \text{Var}(\hat\theta)_{\text{SS}} \simeq\frac{1-\eta^2+2 \eta^2 \sum_{k=1}^\infty \sinh(\sigma^2 c^{k})}{N\eta^2}.
\end{equation}
Note that for the uncorrelated dephasing case $c=0$, we recover the well-known result that
\begin{equation}
\label{eq:uncorbound}
\textrm{Var}(\hat{\theta})\simeq\frac{1-\eta^2}{N \eta^2},
\end{equation}
which saturates the fundamental bound \cite{Escher2011, Demkowicz-Dobrzanski2012}, proving that the strategy is asymptotically optimal.
Similarly, we can compute the variance for the product probe state by setting $\langle\sigma_y^{(i)}\sigma_y^{(j)}\rangle=0$ instead to obtain:
\begin{align}
\text{Var}(\hat\theta)_\text{product} &\simeq\frac{1+2 \eta^2 \sum_{k=1}^\infty \sinh(\sigma^2 c^{k})}{N\eta^2}\\
&=\text{Var}(\hat\theta)_\text{SS}+\frac{1}{N}.\nonumber
\end{align}
The improvement due to squeezing by an additive factor $1/N$ can be generalized for other estimation schemes and noise correlations, and we refer the reader to Appendix~\ref{app:higher order} for the general proof.

In the next section, we will provide a bound based on the classical simulation idea that will allow us to see if there is space for improvement over the spin-squeezed state protocol 
in the presence of correlations.  

%and therefore we can partially perform the summation needed for $\text{Var}(J_y)$:
%\begin{align}
%    \sum_{i,j=1}^N\langle\sigma_y^{(i)}\sigma_y^{(j)}\rangle&=\left(N+\sum_{k=2}^{N-1}2(N-k+1)\langle\sigma_y^{(1)}\sigma_y^{(k)}\rangle\right)\simeq\nonumber\\&\simeq N\left(1+2\sum_{k=2}^\infty\langle \sin\varphi_1\sin\varphi_k\rangle-\eta^2\right),
%\end{align}
%yielding asymptotic estimation precision:
%\begin{equation}
%\label{eq:lower}
%    \text{Var}(\hat\theta)=\frac{1-\eta^2+2\sum_{k=2}^\infty\langle \sin\varphi_1\sin\varphi_k\rangle}{N\eta^2}.
%\end{equation}
%For AR(1) noise \eqref{eq:gaussian} the second moments are:
%\begin{align}
    % \langle\cos\varphi_k\rangle=e^{-\sigma^2/2}\\
%    \langle\sin\varphi_i\sin\varphi_j\rangle&=e^{-\sigma^2}\sinh\left(\sigma^2c^{|i-j|}\right),\\
 %   \langle\cos\varphi_i\cos\varphi_j\rangle&=e^{-\sigma^2}\cosh\left(\sigma^2c^{|i-j|}\right),
% ,\quad D=-\log\eta^2,
%\end{align}
%and unfortunately we cannot express bound \eqref{eq:lower} in a closed form, although for a fixed $c,\sigma$ we can easily compute the sum numerically.

% We obtained upper and lower bound on asymptotic QFI for general Markovian dephasing:
% \begin{align}
% \label{eq:upper}
%     \lim_{N\to\infty}\frac{\mathcal{F}_Q^{(N)}}{N}&\leq \mathcal{F}_{cl}(\varphi_2,\varphi_1)-\mathcal{F}_{cl}(\varphi_1),\\
% \label{eq:lower}
%     \lim_{N\to\infty}\frac{\mathcal{F}_Q^{(N)}}{N}&\geq\frac{\eta^2}{1-\eta^2+2\sum_{k=2}^\infty\langle \sin\varphi_1\sin\varphi_k\rangle}.
% \end{align}

\section{Classical simulation bound}
\label{sec:CS bounds}
%\sk{[Shorter title:Classical simulation bounds]}
To assess the precision of estimation, we now focus on deriving upper bounds on QFI with correlated noise using the \emph{classical simulation (CS) method} \cite{matsumoto2010metricquantumchannelspaces,Demkowicz-Dobrzanski2012}. We call a parametrized quantum channel classically simulable if it admits a decomposition into a convex mixture of  $\theta$-independent channels with $\theta$-dependent classical weights
\begin{equation}
    \Lambda_{\theta}(\rho) = \int \text{d}p_\theta(x)\tilde\Lambda_x(\rho).
\end{equation}
This decomposition allows us to split the estimation error into quantum error---due to the indistinguishability of channels $\{\tilde\Lambda_x\}$, and a classical error---caused by nondeterministic classical mixing with probability distribution $p_\theta(x)$. By neglecting quantum error, we can bound the channel QFI (optimize QFI over all possible input states) by the classical Fisher information (CFI) of the distribution $p_\theta(x)$, 
\begin{equation}
\mathcal{F}_Q(\Lambda_\theta) \leq \mathcal{F}_{\textrm{cl}}[p_\theta(x)] = \int \textrm{d}x\, \frac{1}{p_\theta(x)} \left(\frac{\textrm{d} p_\theta(x)}{\textrm{d}\theta}\right)^2.
\end{equation}
%\sk{Maybe explicit formula here: $\mathcal{F}(\Lambda_\theta) \le F(p_\theta(x))$} 
This bound can be easily applied for model \eqref{eq:N-partite channel} 
if we rewrite the action of the channel as:
\begin{equation}
   \Lambda^{(N)}_\theta( \cdot )=\int\text{d}p_{\theta}(\vec{\xi})\left(\bigotimes_{n=1}^N U_{\xi_n}\right) (\cdot )\left(\bigotimes_{n=1}^NU_{\xi_n}\right)^\dagger,
\end{equation}
where
\begin{equation}
p_{\theta}(\vec{\xi}) = p(\vec{\theta} - \vec{\xi}),
\end{equation}
and $\vec{\theta}=[\theta,\dots,\theta]^T$.
We can now compute the CFI for the classical Gaussian statistical model $p_\theta(\vec{\xi})$, making use of the fact that it represents a Markovian process, we can express the CFI in terms of CFIs of the marginal distribution of neighboring phases and single-phase distribution \cite{Radaelli_2023, amoros2021noisy}:% \sk{[what is $\simeq$ here? I know it was discussed earlier, still it's not clear for me in this context]}
\begin{align}
\label{eq:noise bound}
    \mathcal{F}_{cl}(\vec{\xi})&=(N-1)\mathcal{F}_{cl}(\xi_2,\xi_1)-(N-2)\mathcal{F}_{cl}(\xi_1)\nonumber\simeq\\&\simeq N[\mathcal{F}_{cl}(\xi_2,\xi_1)-\mathcal{F}_{cl}(\xi_1)].
\end{align}
For our model we get
\begin{align}
    \mathcal{F}_{cl}(\varphi_2,\varphi_1)=\frac{2}{1+c}\sigma^{-2},\quad\mathcal{F}_{cl}(\varphi_1)=\sigma^{-2},
    % \mathcal{F}_{cl}(\vec{\varphi})=\sum_{i,j=1}^N(\Sigma^{-1})_{ij}=
\end{align}
which finally leads to the CS bound \cite{macieszczak2014upperboundsquantumfisher}:
\begin{equation}
\label{eq:upper}
    \mathcal{F}_Q^{(N)}\leq N\sigma^{-2}\frac{1-c}{1+c}.
\end{equation}
Unfortunately, this bound is not tight, even for the uncorrelated case $c=0$---comparing with \eqref{eq:uncorbound}, we see that
the bound on QFI should be $F_Q^{(N)} \leq \frac{N\eta^2}{1-\eta^2} = N/(e^{\sigma^2}-1)$, which is strictly  tighter than $N/\sigma^2$. 

\begin{figure}[t]
    \includegraphics[width=\linewidth]{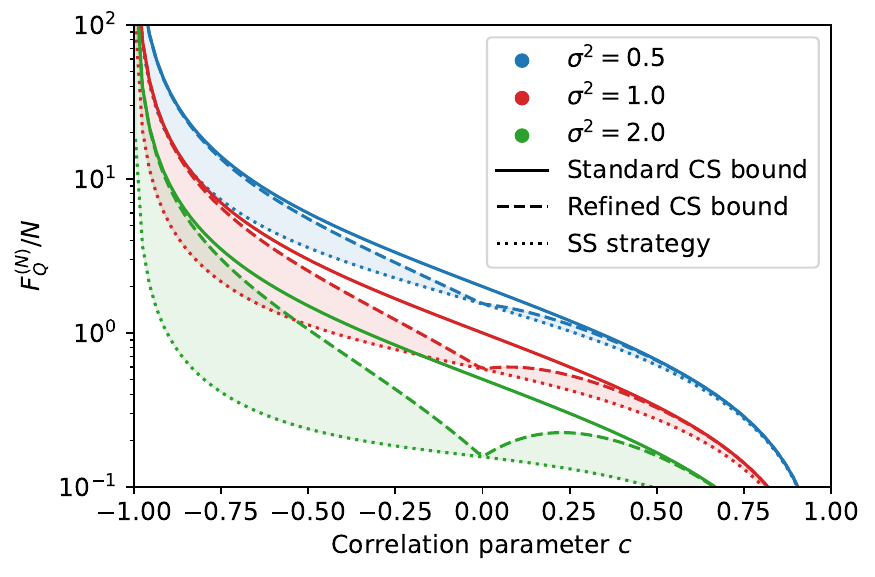}
    \caption{Comparison of standard  CS bound \eqref{eq:upper}---solid lines---and refined CS bound \eqref{eq:new CS bound}---dashed lines---to precision of SS probe state strategy with $J_y$ measurement \eqref{eq:lower}---dotted lines---for  Markovian Gaussian noise model \eqref{eq:gaussian}. The actual QFI lies inside the shaded regions. We see that the refined bound is tight at $c=0$ and is significantly tighter than the standard bound for weakly correlated noise. Moreover, for strong positive correlations, both bounds and SS strategy yield similar FI, coinciding in the limit $c\to1$.}
    \label{fig:old_bounds}
\end{figure}

To solve this problem, we propose a tighter, refined bound, based on decomposition of the Gaussian noise into uncorrelated and correlated parts: 
%[now equation (40) and maybe (36)], where [now equations 45, 46], and consequently [now equation (47), I'm not sure if we want to write it with variance or with FI], see appendix... for more details. The advantage of this decomposition comes from the fact that the uncorrelated part of the bound is tight, as it is equivalent to (15). This means that the resulting bound will be the best in a small-$c$ limit, in which uncorrelated part dominates in (40), see Fig...      \skdel{For Gaussian noise it is convenient to split the noise into maximally noised uncorrelated part---for which we can use the tight bound \eqref{eq:uncorrelated bound}---and complementary correlated part:}
%\sk{This should go to appendix (up to [40]):}
\begin{align}
    \mathcal{N}(\theta,\Sigma)&\sim\mathcal{N}(q\theta,\lambda\mathds{1})+\mathcal{N}((1-q)\theta,\Sigma-\lambda\mathds{1}),\\
    \mathcal{F}_Q^{(N)}&\leq q^2\mathcal{F}_{\small\text{uncorrelated}}+(1-q)^2\mathcal{F}_{\small\text{correlated}},\\
\mathcal{F}_{\small\text{uncorrelated}}&=\left.\mathcal{F}_Q^{(N)}\right|_{c=0, \eta=\eta(\lambda)}=\frac{N\eta(\lambda)^2}{1-\eta(\lambda)^2},\\
    \mathcal{F}_{\small\text{correlated}}&=\mathcal{F}_\text{cl}(\mathcal{N}(\theta,\Sigma-\lambda\mathds{1})),
\end{align}
where $\lambda$ is the smallest eigenvalue of $\Sigma$, so that $\Sigma-\lambda\mathds{1}$ is semi-definite positive, therefore a valid covariance matrix, $\eta(\lambda)^2=e^{-\lambda}$ is the strength of uncorrelated noise and $q\in[0,1]$ controls splitting of $\theta$. Assuming $c \geq 0$, and optimizing over $q$, we obtain:
\begin{equation}
    \label{eq:harmonic}
    \mathcal{F}_Q^{(N)}\leq(\mathcal{F}_{\small\text{uncorrelated}}^{-1}+\mathcal{F}_{\small\text{correlated}}^{-1})^{-1},
\end{equation}
where (see Appendix~\ref{app:CS bound} for more details)
\begin{align}
    \mathcal{F}_{\small\text{uncorrelated}}&=\frac{N}{e^\lambda-1}=\frac{N}{\exp\left(\sigma^2\frac{1-c}{1+c}\right)-1},\\
    \label{eq:spectral sum}\mathcal{F}_{\small\text{correlated}}&%=\sum_{i,j,k=1}^N\frac{V_{ik}V_{jk}}{D_{kk}-\lambda}
    \simeq N\frac{1-c^2}{4c\sigma^2}.
    % \\\nonumber&=\frac{2}{N\sigma^2}\sum_{i,j,k=1}^N\frac{\sin\left({ik}/{N}\pi\right)\sin\left({jk}/{N}\pi\right)}{\frac{1-c^2}{1+c^2-2c\cos(k\pi/N)}-\frac{1-c}{1+c}}
\end{align}
%we refer to appendix \ref{app:CS bound} for detailed derivation.
%\sk{This can also go to appendix}
For $c<0$, above method gives divergent $\mathcal{F}_{\small\text{correlated}}=\infty$, so the bound \eqref{eq:harmonic} simplifies to $\mathcal{F}_Q^{(N)}\leq N/(e^\lambda-1)$, where $\lambda=\sigma^2\frac{1+c}{1-c}$. Finally, the refined CS bound reads:
% \begin{equation}
% \label{eq:new CS bound}
%     \frac{F_Q^{(N)}}{N} \leq \left\{\begin{matrix} \left[\exp\left(\sigma^2\frac{1-c}{1+c}\right)+\sigma^2\frac{4c}{1-c^2}-1 \right]^{-1}&\text{when $c>0$,}\\\left[\exp\left(\sigma^2\frac{1+c}{1-c}\right)-1 \right]^{-1}&\text{when $c\leq0$.}\end{matrix}\right.
% \end{equation}
\begin{equation}
\label{eq:new CS bound}
\frac{F_Q^{(N)}}{N} \leq
\begin{cases}
\left[\exp\left(\sigma^2\frac{1-c}{1+c}\right)+\sigma^2\frac{4c}{1-c^2}-1 \right]^{-1}
\quad c>0, \\[6pt]
\left[\exp\left(\sigma^2\frac{1+c}{1-c}\right)-1 \right]^{-1}
\quad c\leq 0.
\end{cases}
\end{equation}
%\sk{To appendix} Let us note that using this method we can derive CS bound to general OU driven process, without assumption $\alpha\delta\approx 0$, as the correction term is proportional to identity matrix.
Comparison of CS bounds with respect to known achievable precision \eqref{eq:lower} is presented in Fig.~\ref{fig:old_bounds}. We observe significant improvement of the refined CS bound \eqref{eq:new CS bound} over the standard bound \eqref{eq:upper} for weak correlations. Moreover, the analytical form of the bound enables a clear distinction between correlated and uncorrelated noise scenarios, thereby providing valuable insights into the effects of noise correlations.

%The above analysis assumes number of qubits $N$ is the limited resource, we can also try to use the bounds for time-continues OU model with total time $T=N\delta$ being the limited resource and $\omega_0$ being the estimated parameter. By using \sk{[So we use looser bound because it is saturable anyway for this case? I'm not sure if I understand it]} looser bound \eqref{eq:upper} and substituting values from equations (\ref{eq:OU1},\ref{eq:OU2},\ref{eq:OU3}) we obtain:
%\begin{equation}
%    \text{Var}(\hat\omega)\geq \frac{\sigma^2}{N}\frac{1+c}{1-c}\left|\frac{\partial\omega}{\partial\theta}\right|^2=\frac{\delta\beta^2}{2\alpha T}\frac{1+e^{-\alpha\delta}}{1-e^{-\alpha\delta}}\geq\frac{\beta^2}{\alpha^2}T^{-1},
%\end{equation}
%where the optimal estimation is achieved for $\delta\to 0$ and $N\to\infty$. Moreover, the bound can be saturated by estimator \eqref{eq:old_estimator} for product probe state, which can be seen by analysing Eq. \eqref{eq:lower}. Although this limit is not physical, as it requires qubit manipulations in infinitesimal time, it shows that performing more measurements can be beneficial, even if the outcomes are highly correlated.

We see that for positive correlations $c \geq0$, spin-squeezed states perform extremely well, and the bound indicates that there is not much space for further improvement. On the other hand, we see that for negative correlations, $c < 0$, the gap between the spin-squeezed state protocol and the bound is larger, giving hope for the existence of better protocols.
% \sk{In (39), do you mean that $\mathcal{F}$ is the FI of classical simulation, which coincides with QFI for uncorrelated case? Maybe it's better to just say that $\mathcal{F}$ is classical simulation FI. Also, do we need this Fourier?}

Generally, we can derive the refined CS bound for any dephasing model, without assuming Gaussianity or the Markov property. If we find decomposition of noise into several $N$-variate random variables $\mathcal{R}_k$
\begin{align}
\label{eq:spliting}
    &\vec\varphi\sim \sum_{k=1}^n \vec{\mathcal{R}}_k, &\mathcal{F}_Q^{(N)}\leq\left(\sum_{k=1}^n\mathcal{F}(\vec{\theta}+\vec{\mathcal{R}}_k)^{-1}\right)^{-1}
\end{align}
where $\mathcal{F}$ is the FI of classical simulation, which is equal to QFI for the uncorrelated variables. On the level of probability distributions condition \eqref{eq:spliting} can be written as
\begin{equation}
    p(\vec\varphi)=p(\vec{\mathcal{R}}_1)\ast p(\vec{\mathcal{R}}_2)\ast\dots\ast p(\vec{\mathcal{R}}_n)
\end{equation}
where $\ast$ is the convolution operator. For many cases, it is easy to decompose $p(\vec\varphi)$ in this way by working on the Fourier transform of distributions, where convolution is mapped to multiplication. For Gaussian variables, this condition is simply that the sum of covariance matrices of $\vec{\mathcal{R}}_k$ is $\Sigma$.

\section{Optimized protocols using tensor network approach}
\label{sec:tensor network}
There are mainly two well-known approaches for computing QFI numerically. The first one, the minimization over purifications, has some serious drawbacks \cite{Kurdzialek2024}; therefore, we follow the \textit{iterative see-saw (ISS)} approach, which is far more convenient.
% depends on the fact that the QFI for any mixed state may be equivalently expressed as a minimization of QFI for all of its admissible purifications $|\psi_{\theta}\rangle$. 
% Additionally, this method can be cast as a \textbf{semi-definite programming (SDP)} problem to find the optimal input probe state for the estimation problem. 
% However, this approach has some serious drawbacks \cite{Kurdzialek2024}. On the other hand, the \textit{iterative see-saw (ISS)} approach is far more convenient compared to the previous one.
% \skdel{Apart from QFI optimization, ISS can be employed to deal with even Bayesian and minimax analysis, also it works efficiently for the Channel extension method, because one can handle the ancillary system separately, which is not possible in the case of MOP.} 
% In this section, we use the ISS approach to find out the optimal probe and the fundamental bounds for the optimal estimation strategy. 
% \subsubsection*{Iterative see-saw approach}
% As mentioned earlier, ISS can handle the inclusion of the ancillary system separately. 
% To define the ISS approach, we write $\Lambda_{\theta}$ instead of $\Lambda_{\theta}\otimes \mathcal{I}$.
Let us consider the \textit{pre-QFI function}
\begin{equation}
    \mathcal{F}(\rho,L) = 2 \mathrm{Tr}(\dot{\rho}_{\theta}L) - \mathrm{Tr}({\rho_{\theta}L^{2}}),\quad \rho_\theta = \Lambda_\theta(\rho),
    \label{pre-qfi}
\end{equation}
where $L$ is a Hermitian operator. The quantum Fisher information optimized over all input states $\rho$ can be obtained as $\mathcal{F}_{Q} = \max_{\rho, L} \mathcal{F}(\rho, L)$.
% \sk{[We do not need this]} 
% Note that, Eq. \eqref{pre-qfi} can also be rewritten as

% \begin{equation}
    % \mathcal{F}(\rho,L) = 2 \mathrm{Tr}(\rho \dot{\Lambda}^{*}_{\theta}(L)) - \mathrm{Tr}({\rho \Lambda^{*}_{\theta}(L^{2})}).
    % \label{pre-qfi_dual}
% \end{equation}
% Here $\Lambda_{\theta}^{*}$ denotes the dual map of $\Lambda_{\theta}$. 

\begin{figure*}[t]
    \includegraphics[width=\textwidth]{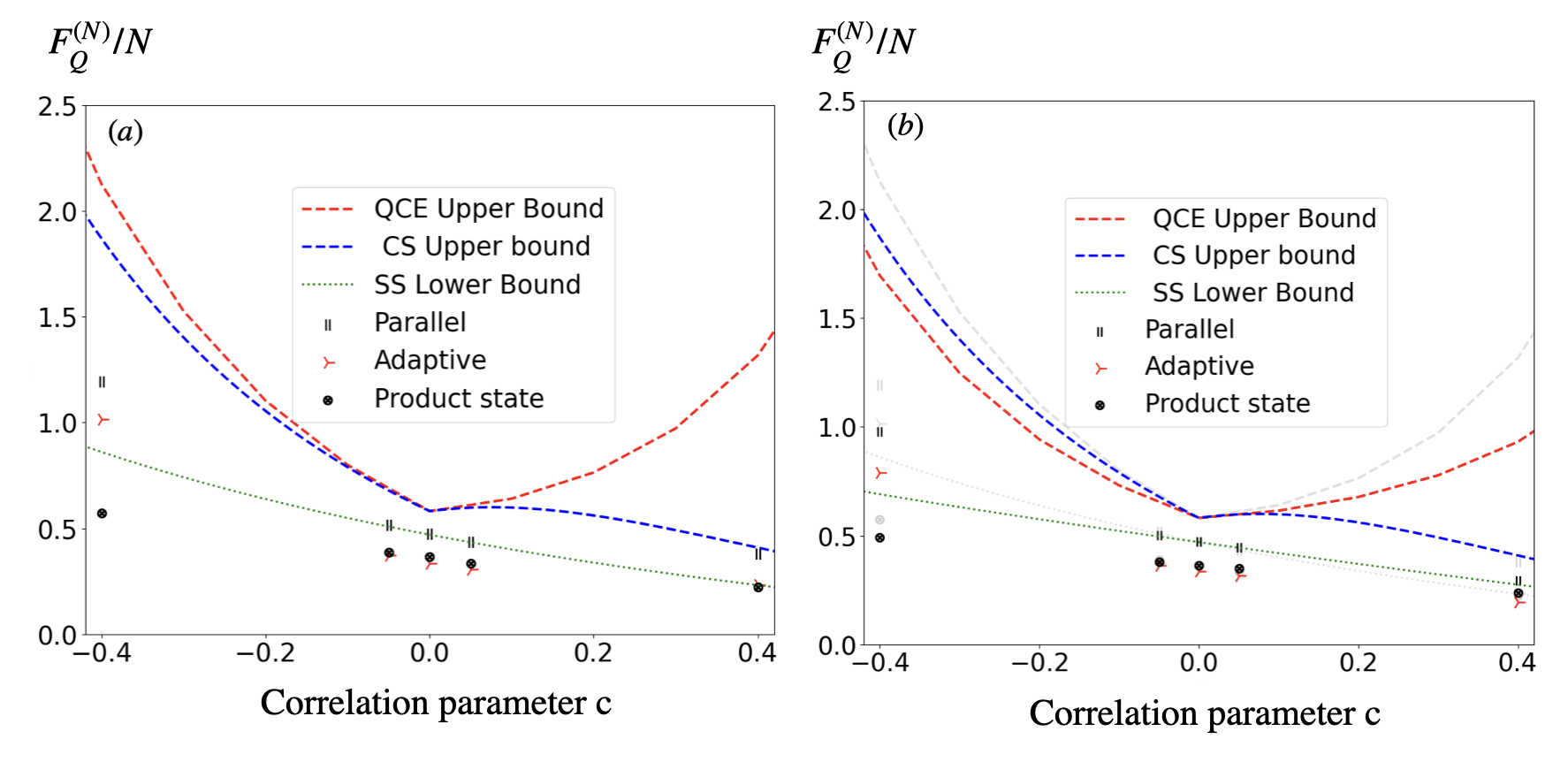}
    \caption{\justifying The dashed lines indicate the asymptotic upper bounds of the quantum Fisher information per channel, $\lim_{N\rightarrow \infty}\mathcal{F}_{Q}^{(N)}/N$, as a function of the correlation parameter $c$. \textbf{(a)} The left panel contains the binary phase, i.e., $M = 2$ and \textbf{(b)} the right panel demonstrates $M=4$. The red curve represents the upper bound obtained from the QCE method with the Rouwenhorst approximation, and the blue curve is the tighter upper bound obtained by the classical simulation bound \eqref{eq:new CS bound} for the actual Gaussian noise model. The dotted green line denotes the lower bound for the approximate model with an SS state as the initial probe state at $N = 30$. The scattered points correspond to the computed values of $\mathcal{F}_{Q}^{(N)}/N$ with $N = 30$ for the discretized model, obtained via tensor network optimization. For the parallel strategy, the bond dimension of the input state is considered to be $4$ and $1$, which correspond to entangled (marked as $||$) and product states (marked as $\bullet$), respectively. For the adaptive strategy, we consider the ancilla dimension to be $3$. We also plot results for $M=2$ on the right panel with grayed lines to better appreciate the effect of more refined phase discretization. Here $\sigma^2 = 1$.}
    \label{fig:phase_4}
\end{figure*}
The double maximization of the ISS protocol is an iterative process that goes as follows. Firstly, perform the maximization of $\mathcal{F}(\rho,L)$ over $L$ for an arbitrary input state (educated guess) $ \rho =\rho^{0}$, to obtain $L^{0}$. Then for the aforementioned $L^{0}$, again maximize the function with respect to $\rho$ to obtain $\rho^{1}$. One needs to repeat this procedure until the function $\mathcal{F}(\rho,L)$ converges to some $\mathcal{F}(\rho^{i},L^{i})$.
% This double maximization of Eq. \eqref{pre-qfi} over $\rho$ and $L$ gives yields the quantum Fisher information; hence, $\mathcal{F}_{Q} = \max_{\rho, L} \mathcal{F}(\rho, L)$. Note that this convergence is guaranteed in generic cases \cite{}. 
Both steps of optimization boil down to solving a simple convex optimization problem, either an unconstrained quadratic in the case of optimization with respect to $L$, or a semi-definite programming problem with $\rho\geq 0,\mathrm{Tr}(\rho)=1$ constraints in the latter case.

For a single-channel application, the aforementioned method can accurately determine the ideal probe condition for any metrological protocol. For multiple uses of the channel, if all of them are uncorrelated, one can use the ISS method with $\Lambda_{\theta}$ substituted by $\Lambda_{\theta}^{\otimes N}$. The mathematical tool used to handle the adaptive protocol is the quantum comb \cite{Chiribella2008a, Chiribella2009}. 
% However, if quantum channel can be used multiple times in the protocol, one can outperform the classical limit for some parameter estimation problem. ISS can also be efficiently generalised to the many channel uses within the most general, adaptive estimation scheme.
% which provides a sequential or parallel method within a specified limit, 
For the correlated channel estimation,  the $N-$partite quantum channel $\Lambda_{\theta}^{(N)}$ does not follow product structure $\Lambda_{\theta}^{(N)} \ne \Lambda_{\theta}^{\otimes N}$. The optimization over combs becomes inefficient for large $N$, as the size of the optimized comb grows exponentially with $N$. To avoid this, one can efficiently represent a quantum comb with a tensor network \cite{kurdzialek2025}.

To model classical Markovian noise we introduce auxillary quantum environment with orthogonal basis indexed by Markov chain state-space $\{|\phi\rangle\}$ and a quantum channel defined with Kraus operators indexed by pairs of basis vectors:
\begin{equation}
\label{eq:kraus}
    K({\phi_{\tiny\text{IN}},\phi_{\tiny\text{OUT}}})=\sqrt{p({\phi_{\tiny\text{OUT}}|\phi_{\tiny\text{IN}}})}U_{\phi_{\tiny\text{OUT}}}\otimes|\phi_{\tiny\text{OUT}}\rangle\langle\phi_{\tiny\text{IN}}|,
\end{equation}
where $p({\phi_{\tiny\text{OUT}}|\phi_{\tiny\text{IN}}})$ is the transition probability distribution of the Markov chain. The state of the auxiliary system transfers the state of the Markov chain, while the probe system experiences random unitary rotation with corresponding probabilities. 
% For the AR(1) model the state space is $\mathbb{R}$, and therefore the auxillary space is inifinite-dimensional. 
For implementing this approach numerically, we need a discretized version of our continuous noise model, which closely approximates the stochastic correlations while benefiting from a finite state-space $\{\phi_k\}$ with $k=1, \ldots, M$. From a broad family of potential discretizations, we choose the \textit{Rouwenhorst process} \cite{KOPECKY2010701}.
% which perfectly recreates the autocorrelation function of AR(1) and has 
The transition probability distribution is defined as coefficients of a polynomial:
\begin{align}
\label{eq:transition}
    &\sum_{i=1}^Mp^{(c)}(\phi_i|\phi_j)t^{i-1}=2^{1-M}\\& \left[\left({1+c}\right) + \left({1-c}\right)t\right]^{M-j}\left[\left({1-c}\right) + \left({1+c}\right)t\right]^{j-1}.\nonumber
\end{align}
These transition probabilities satisfy
\begin{equation}
\label{eq:monoid}
    \sum_{k=1}^Mp^{(c_1)}(\phi_i|\phi_k)p^{(c_2)}(\phi_k|\phi_j)=p^{(c_1\cdot c_2)}(\phi_i|\phi_j).
\end{equation}
The condition \eqref{eq:monoid} guarantees that if $\langle\varphi_k\rangle,\text{Var}(\varphi_k)$ and $\langle\varphi_k\varphi_{k+1}\rangle$ for the Rouwenhorst process and the continuous Gaussian process are equal, then the whole covariance matrices are identical. The stationary distribution of the Rouwenhorst process is a binomial distribution $p(\phi_k)=2^{1-M}\binom{M-1}{k-1}$, and therefore, due to De Moivre-Laplace theorem the correct choice of states $\{\phi_k\}$ are equally spaced in an interval symmetric around $0$ of width dependent on the noise strength---in our case we numerically find the width such that
% such that both AR(1) and Rouwenhorst method have identical values of
$\eta=\langle\cos\varphi\rangle$ for stationary distributed $\varphi$. The equality of $\langle\varphi_k\varphi_{k+1}\rangle$ holds when the correlation parameter $c$ is identical in both the Gaussian and Rouwenhorst processes.
% \textcolor{blue}{With this choice of state-space, the transition probabilities \eqref{eq:transition} are uniquely determined by the assumption that the discretised process has exactly the same covariance matrix as the original Gaussian noise.}
With matched first and second moments, we can approximate the Gaussian noise with the derived discretisation, and this approximation is especially good for moderate values of $c$ or large $M$.
% In the limit $M\to\infty$ the state space becomes dense in the $\mathbb{R}$ and the probability distribution distributions defining Rouwenhorst process approaching corresponding AR(1) distributions---in case of the stationary distribution this is a well known de Moivre-Laplace theorem.

Using the obtained Kraus operators \eqref{eq:kraus}, we determine the optimal schemes for both parallel and adaptive scenarios using the functions \texttt{iss\_tnet\_adaptive\_qfi} and \texttt{iss\_tnet\_parallel\_qfi}, from the QMetro++ Python optimization package \cite{dulian_package}. These functions are efficient even for a moderately large ($N \approx 30$) number of channels thanks to tensor network methods. These tensor network decompositions assume a limited available memory size for adaptive protocols and limited bond dimension of the input entangled state for parallel protocols.
% \skdel{Note that the parallel strategy can be derived from the most general adaptive strategy when the control operations become swap operations \cite{dulian_package}.} 
The results for $M=2$ and $M=4$ are shown in Fig.~\ref{fig:phase_4}.
% The values of the binary phases $\epsilon$ come out to be $(-0.919,0.919)$, which obey Eq. \ref{phase_match}.
The bond dimension of MPS in parallel optimization was fixed at $4$ and the memory size in adaptive optimization was set to $3$, which allowed for efficient numerical optimization in reasonable time up to the number of channels used $N\approx 30$. 

%One important point is the 
%It turns out that for these choices of parameters, parallel strategies work better than adaptive.
% \skdel{It turns out, to saturate the bound asymptotically, the parallel strategy works better than the adaptive one.}

Additionally, we compute fundamental upper bounds for QFI using the method \texttt{ad\_asym\_bounds\_correlated} from the QMetro++ package. These bounds are based on the \textit{quantum comb extension} (QCE) technique, in which the QFI of a more powerful strategy, in which control comb may affect normally inaccessible environments, is used as an upper bound for the QFI of normal adaptive schemes. The bounds are valid both for parallel and adaptive strategies with unlimited memory and MPS bond dimension, see \cite{kurdzialek2025} for more details.  
% It is found that, in the asymptotic limit, the parallel strategy outperforms the adaptive one in saturating the ultimate bound. 
For both positive and negative correlations, the bond dimensions of the probe state and the corresponding \textcolor{blue}{SLD} operators are equal to $4$,
%\sk{Im not sure if it makes sense to compare with adaptive strategy, for which bond dimension has completely different meaning.} \sg{I agree with you, but probably here we are trying to convey the message that with this ancilla dimension adaptive strategy performs worse than parallel with bond dimension 4} 
exceeding that of the adaptive strategy. Note that the bounds are asymptotically valid for $N\rightarrow \infty$, which explains a gap between the bounds and the performance of the actual protocols \textcolor{blue}{for} $c=0$, which will close if one increases $N$ further.

A detailed numerical investigation indicates that for both moderately positive, negative and also for uncorrelated scenarios, (i.e., approximately, $|c| \le 0.5$) the difference in the computed Fisher information between discretizations with $M = 4$ and $M = 5$ phases is negligible for all practical purposes (see Appendix~\ref{app:numerics}), % \textcolor{blue}{A detailed numerical investigation indicates that for both moderately positive and negative correlations ($c = \pm 0.4$), i.e., the correlation is neither very strong ($c = \pm 1$) nor small enough to be regarded as negligible, as well as for the uncorrelated case ($c = \pm 0.0$), the difference in the computed Fisher information between discretizations with $M = 4$ and $M = 5$ phases is negligible for all practical purposes (see Appendix~\ref{app:numerics})}, 
indicating that $M=4$ discrete phases are sufficient to faithfully approximate the continuous noise. Although increasing the value of $M$ yields a finer approximation of the continuous noise, the improvement is not substantial enough to affect the determination of the optimal strategy for the protocol.
Therefore, to ensure computational efficiency while maintaining sufficient accuracy, we adopted $M = 4$ throughout the numerical simulations.

Our systematic analysis reveals that although the CS method is generally considered weaker than the QCE approach \cite{kurdzialek2025} for estimating the upper bound of the Fisher information, in the presence of positive correlations, the CS method provides a tighter bound. Notably, both bounds coincide near $c = 0$. We determine the optimal strategies that saturate these bounds using SS states and matrix product states with a bond dimension of $4$. As shown in Fig.~\ref{fig:phase_4}, the parallel strategy proves to be more efficient than the adaptive one (with the assumed ancillary space dimension) in reaching the asymptotic limit---it appears that in order to reach the asymptotic performance in the adaptive paradigm, the size of the ancillary system would need to grow with the number of channel uses and hence it is much more efficient to look for optimal strategies using parallel strategies involving MPS states. 

Indeed, for both positive and negative correlations, MPS as probe states outperform the SS state strategy, with the advantage being particularly pronounced for $c < 0$. 
% As discussed earlier, negatively correlated channels exhibit superior performance compared to those with positive correlations, and this trend persists even for a moderate number of channel uses (see Appendix~\ref{app:numerics}). 
%In both the uncorrelated and positively correlated regimes, MPS with a bond dimension of $4$ perform slightly better than SS states. 
Increasing the bond dimension is expected to further enhance performance; however, for the uncorrelated and positively correlated case $c > 0$, this advantage is expected to fade away asymptotically when $N\rightarrow \infty$. In contrast, for $c < 0$, the advantage of the MPS approach over the SS strategy is substantial, and already for finite $N \approx 30$ provides a ratio $F_Q/N$ larger than the asymptotic ratio $\lim_{N \rightarrow \infty} F_Q/N$ achievable with SS strategy 
%\rdd{Srijon double check this?  I also modified some text above, the claim that MPS is better for positive c, was to strong in my opinion}. 

Moreover, one might
wonder if including measurements of higher-order observables in the SS strategy could offer comparable improvement. We have verified that this advantage may only be observed in strong noise and strong negative correlation regimes and will not affect the performance for the moderate noise strength $\sigma^2=1$ discussed above—see Appendix \ref{app:higher order} for detailed discussion.

% \begin{figure}[H]
%     %\centering
%     \includegraphics[width=\linewidth]{Moderate_comparison_25_10_2025.png}
%     \caption{Quantum Fisher information per channel $\mathcal{F}_{Q}^{(N)}/N$ as a function of the number of channel uses $N$. The solid lines represent the upper bound of the Fisher information computed with a block size of three. The dashed and dotted lines correspond to $\mathcal{F}_{Q}^{(N)}/N$ values for the initial probe states, which are matrix product states obtained via tensor-network optimization and spin-squeezed states, respectively. Red, green, and blue curves denote correlation parameters $c = 0.0$, $c = -0.4$, and $c = 0.4$, respectively. The number of phases is fixed at four.}

%     \label{fig:positive-negative comparison}
% \end{figure}

\section{Discussion}
\label{sec:conclusion}

In this work, we have investigated a metrological frequency estimation protocol in which phase-encoding quantum channels exhibit noise correlations.

The general tools to optimize metrological protocols and derive fundamental bounds
for correlated cases have been proposed in Ref.~\cite{Kurdzialek2024} and Ref.~\cite{kurdzialek2025}, respectively.
The toy model involving binary correlated dephasing, for which random rotations angles
$\varphi_n$ take only two possible values, has been studied in both aforementioned
references as an illustration of the introduced techniques.

However, the generalization to more realistic models in which $\varphi_n$ are given by a continuous-valued stochastic process remains a non-trivial task.
The methods derived in Refs.\cite{Kurdzialek2024, kurdzialek2025} cannot be directly applied in this case, as
doing so would require the use of intractable combs with infinite memory to properly
model noise correlations.

In this work, we overcome this difficulty and, for the first time, comprehensively study metrological scenarios involving correlated dephasing governed by a realistic Gaussian Markov process.

To derive fundamental metrological bounds, we use the CS technique, which is
applicable directly to continuous-variable processes.
We generalize a known uncorrelated bound \cite{Demkowicz-Dobrzanski2012} to cases involving correlations.
We further tighten the obtained bound by incorporating the novel idea of dividing
the noise into correlated and uncorrelated parts and using an appropriate classical
simulation for each part independently.

Furthermore, we analyze the performance of squeezing-based metrological protocols—such protocols have so far been studied for uncorrelated noise only. We show that squeezing remains a valuable metrological resource also in the presence of correlations. An interesting difference between uncorrelated and correlated cases is that higher spin moments may be an additional source of information in the latter case, as discussed in Appendix~\ref{app:higher order}.

Finally, we show how to approximate a continuous noise model with a discretized model using the Rouwenhorst method. This allows to apply the techniques developed in Refs.\cite{Kurdzialek2024, kurdzialek2025}
for a comparison.

%By employing the classical simulation method, we derived a new and tighter upper bound on the quantum Fisher information for such correlated channels. Our comparative analysis reveals that this bound surpasses the previously established limit obtained via the QCE method, particularly for positively correlated channels. To handle the correlated Gaussian noise model, we discretized the dynamics using the Rouwenhorst method and identified the optimal scheme for saturating the derived bounds through tensor network optimization of both the input probe state and the measurement strategy. Our results demonstrate that  MPS with finite and small bond dimensions outperform spin-squeezed states when used as probe states within the protocol. Furthermore, we found that even the most general adaptive strategies, when constrained by finite ancilla dimensions, underperform relative to MPS-based parallel strategies with bond dimension of comparable size as the ancilla dimension in the adaptive strategy. These findings collectively deepen our understanding of metrological optimization in correlated quantum channels and open pathways for more efficient implementations of frequency estimation protocols in realistic noisy environments.

\section{Codes and data availability}
The computer codes for generating numerical data used in this paper are available at \cite{codes}.

\section*{Author Contribution Statement}

S.G. and A.K. contributed equally to this work. All authors contributed to the conceptualization of the research, development of the theoretical framework, analysis of the results, and preparation
of the manuscript. All authors reviewed and
approved the final version.

\section*{ACKNOWLEDGMENTS}
S.G. thanks Arpan Das for fruitful discussions and Piotr Dulian
for the support in running the numerics with the help of the QMetro++ package. S.K. is a recipient of the Foundation for Polish Science START 2025 scholarship. This work was supported by National Science Center (Poland) grant No.2020/37/B/ST2/02134.

% \bibitem{examplecitation}
%   Name Surname,
%   \href{https://doi.org/10.22331/
%         idonotexist}{Quantum
%         \textbf{123}, 123456 (1916).}

% \bibitem{biblatexsubmittingtothearxiv}
%   StackExchange discussion on \href{http://tex.stackexchange.com/questions/26990/biblatex-submitting-to-the-arxiv}{``Biblatex: submitting to the arXiv'' (2017-01-10)}

% \bibitem{arxivpdfoutput}
%   Help article published by the arXiv on \href{https://arxiv.org/help/submit_tex}{``Considerations for TeX Submissions'' (2017-01-10)}

% \bibitem{howtogetdoilinksinbibliography}
%   StackExchange discussion on \href{http://tex.stackexchange.com/questions/3802/how-to-get-doi-links-in-bibliography}{``How to get DOI links in bibliography'' (2016-11-18)}

% \bibitem{automaticallyaddingdoifieldstoahandmadebibliography}
%   StackExchange discussion on \href{http://tex.stackexchange.com/questions/6810/automatically-adding-doi-fields-to-a-hand-made-bibliography}{``Automatically adding DOI fields to a hand-made bibliography'' (2016-11-18)}
% \end{thebibliography}

\appendix

\section{Discretization of a time-continuous frequency estimation model with Ornstein-Uhlenbeck fluctuations}
\label{sec:ou}
We focus on detailed description of experimentally relevant example of time-varying, correlated process, the \emph{Ornstein-Uhlenbeck (OU) model} \cite{OrnsteinUhlenbeck1930,stockton2004robust,petersen2006estimation,amoros2021noisy,amoros2025tracking}, which is constructed with independent Gaussian increments:
\begin{equation}
    \text{d}\omega(t)=\alpha(\omega-\omega(t))\text{d}t+\beta\text{d}W(t),
\end{equation}
where $\alpha,\beta>0$, $W(t)$ is the Wiener process, and $\omega$ is the mean value. The mean-revertability, critical for estimation protocol, is introduced via drift coefficient $\alpha$, which controls the strength of correlations. By integrating the process from time $t_n=(n-1)\delta t$ to $t_{n+1} = n \delta t$, we obtain:
\begin{align}
    \label{eq:OU1}
    \varphi_n&=\int_{t_n}^{t_{n+1}}\omega(t)\text{d}t,\quad\langle\varphi_n\rangle=\delta t\omega,\\
    % \langle\varphi_n\rangle&=\frac{\omega(t_n)-\omega_0}{\alpha}(1-e^{-\alpha\delta})+\omega_0\delta\approx\omega(t_n)\delta,\\
    % \langle\varphi_{n+1}|\omega(t_n)\rangle=\delta\omega_0+e^{-\alpha\Delta}(\langle\varphi_n|\omega(t_n)\rangle-\delta\omega_0)\\
    \label{eq:OU2}
    \text{Var}(\varphi_n)&=\frac{\beta^2}{\alpha^3}(e^{-\alpha\delta t}+\alpha\delta t-1)\approx\frac{\beta^2}{2\alpha}\delta t^2,\\
    \label{eq:OU3}
    \text{Cov}(\varphi_i,\varphi_j)&=\frac{2\beta^2}{\alpha^3}\sinh^2\left(\frac{\alpha\delta t}{2}\right)e^{-\alpha\delta t|i-j|}\approx\\
    &\approx\frac{\beta^2}{2\alpha}\delta t^2e^{-\alpha\delta t|i-j|},\quad i\neq j,\nonumber
\end{align}
where the approximations are valid for $\alpha\delta t\approx0$---in this regime the covariance matrix is identical to model \eqref{eq:gaussian} for $c=e^{-\alpha\delta t}\approx 1$. 
% We can also compute conditional expectancies of $\varphi_n$ given $\omega(0)$ for this process:
% \begin{align}
%     \langle\varphi_{n}|\omega(0)\rangle=\delta\omega+\frac{e^{-\alpha\delta(n-1)}}{\alpha}(\omega(0)-\omega)(1-e^{-\alpha\delta}).
% \end{align}
% \skdel{Using these results,} instead of analysing time-continuous OU process we can just focus on discrete stochastic process $\varphi_n$, which is in fact known as the \textbf{autoregressive Gaussian model AR(1)} \cite{brockwell1991time}. \skdel{It enjoys being the unique time-homogenous Markov chain described by multivariate Gaussian distribution:} \sk{Using \eqref{} one can show that vector of subsequent phases $\vec \varphi = [\varphi_1, \varphi_2,...]^T$ is normally distributed, }  
% \begin{align}
%     \vec{\varphi}&\sim\mathcal{N}(\theta,\Sigma),\quad\Sigma_{ij}=\sigma^2c^{|i-j|},\\
%     \varphi_{n+1}|\varphi_n&\sim \theta+c(\varphi_n-\theta)+\mathcal{N}\left(0,\frac{\sigma^2}{1-c^2}\right),
% \end{align}
% with $c\in(-1,1)$ being the correlation parameter, $\sigma$ being the noise strength and $\theta=\delta\omega_0$ being the mean value of the process to be estimated. For AR(1) model derived from OU process $c=e^{-\alpha\delta}>0$, but we can extend the model to negative correlations, which might occur from underlying process other than OU. \sk{[ I don't understand next two sentences. Do we see non-markovainity from theses equations? Even if the process is non-markovian, it still typical AR(1), right? Maybe delete next two sentences?]}
Whenever $\alpha\delta\approx0$, $\vec{\varphi}$ does not satisfy the Markov property (the integral is not sufficient statistics for deriving the conditional probability distribution with the Markov property), and the covariance matrix differs from the Markovian model by a factor proportional to the identity matrix. In fact, all results of this paper may be used for this model, either directly (SS strategy and CS bound with changed values of $\eta$ and $\lambda$) or with slight modification (using a hidden Markov model instead of a Markov chain for numerical analysis).

% and $\theta=\delta\omega_0$ being the mean value of the process to be estimated. For AR(1) model derived from OU process $c=e^{-\alpha\delta}>0$, but we can extend the model to negative correlations, which might occur from underlying process other than OU. \sk{[ I don't understand next two sentences. Do we see non-markovainity from theses equations? Even if the process is non-markovian, it still typical AR(1), right? Maybe delete next two sentences?]}Let us note that whenever $\alpha\delta\approx0$ assumption is not satisfied, $\vec{\varphi}$ does not satisfy Markov property, which might be suprising as OU is in fact a Markov process. Although we focus on typical AR(1) process, the results of this paper can be also directly applied to the general case whenever $\vec{\varphi}$ has exponentially vanishing correlations.

Let us also translate the CS bound \eqref{eq:upper} for the OU model estimation with total sensing time $T=N\delta t$ being the limited resource, instead of the number of qubits $N$:
\begin{equation}
   \text{Var}(\hat\omega)\geq \frac{\sigma^2}{N}\frac{1+c}{1-c}\left|\frac{\partial\omega}{\partial\theta}\right|^2=\frac{\delta t\beta^2}{2\alpha T}\frac{1+e^{-\alpha\delta t}}{1-e^{-\alpha\delta t}}\geq\frac{\beta^2}{\alpha^2}T^{-1},
\end{equation}
where the optimal estimation is achieved for $\delta\to 0$ and $N\to\infty$. Moreover, the bound can be saturated by estimator \eqref{eq:old_estimator} for the product probe state. Although this limit is not physical, as it requires qubit manipulations in infinitesimal time, it shows that performing more measurements is beneficial, even if the outcomes are highly correlated.

\section{CFI of correlated part of Markovian Gaussian process}
\label{app:CS bound}
We derive Eq. \eqref{eq:spectral sum} for the CFI of correlated part of Gaussian noise.
As $\mathcal{N}(\vec{\theta},\Sigma-\lambda\mathds{1})$ is no longer a Markov chain, we have to compute its CFI via an explicit result for multivariate Gaussian variables \cite{porat1986computation}:
\begin{equation}
    \mathcal{F}_\text{cl}(\mathcal{N}(\vec\theta,\Sigma-\lambda\mathds{1}))=\sum_{i,j=1}^N(\Sigma-\lambda\mathds{1})^{-1}_{ij},
\end{equation}
which requires inversion of a large matrix. This can be performed for our case, as there exists a known asymptotic eigendecomposition of $\Sigma$ \cite{BOGOYA2016606}:
\begin{align}
    \Sigma&=VDV^T,\quad V_{ij}\simeq\sqrt{\frac{2}{N+1}}\sin\left(\frac{ij}{N+1}\pi\right),\\
    D_{ij}&\simeq\delta_{ij}\sigma^2\frac{1-c^2}{1+c^2-2c\cos\left(\frac{i\pi}{N+1}\right)},\quad\lambda=\sigma^2\frac{1-c}{1+c},
\end{align}
for $c\geq 0$ and
\begin{align}
    V_{ij}&\simeq\sqrt{\frac{2}{N+1}}\sin\left(\frac{ij}{N+1}\pi\right)(-1)^{i+j},\\
    D_{ij}&\simeq\delta_{ij}\sigma^2\frac{1-c^2}{1+c^2-2c\cos\left(\frac{i\pi}{N+1}\right)},\quad\lambda=\sigma^2\frac{1+c}{1-c},
\end{align}
for $c<0$. Let us note that we chose $\lambda$ to be less than the smallest eigenvalue of $\Sigma$, but converging to it in the limit $N\to\infty$, which still provides a valid bound for finite $N$ and does not change the asymptotic bound, but ensures that $\Sigma-\lambda\mathds{1}$ is invertible. Let us assume $c\geq 0$ and use the eigendecomposition to write:
\begin{equation}
    (\Sigma-\lambda\mathds{1})^{-1}=V(D-\lambda\mathds{1})^{-1}V^T,
\end{equation}
and therefore:
\begin{align}
    \mathcal{F}_{\small\text{correlated}}&=\sum_{i,j,k=1}^N\frac{V_{ik}V_{jk}}{D_{kk}-\lambda}\simeq\\&\simeq\frac{2}{N\sigma^2}\sum_{i,j,k=1}^N\frac{\sin\left(\frac{ik}{N+1}\pi\right)\sin\left(\frac{jk}{N+1}\pi\right)}{\frac{1-c^2}{1+c^2-2c\cos(k\pi/(N+1))}-\frac{1-c}{1+c}}\nonumber
\end{align}

% We present the computations of sum appearing in Eq. \eqref{eq:spectral sum}:
% \begin{equation}
% \mathbb{S}\simeq\sum_{i,j,k=1}^N\frac{\sin\left(\frac{ik}{N+1}\pi\right)\sin\left(\frac{jk}{N+1}\pi\right)}{\frac{1-c^2}{1+c^2-2c\cos(k\pi/(N+1))}-\frac{1-c}{1+c}}.
% \end{equation}
First, let us compute the sum:
\begin{align}
\label{eq:sum to integral}
    \sum_{i=1}^N\sin\left(\frac{ik}{N+1}\pi\right)&\simeq N\int_0^1\sin(k\pi x)\text{d}x=\nonumber\\&=\left\{\begin{matrix}\frac{2N}{k\pi} & \text{if $k$ is odd,}\\
    0 & \text{if $k$ is even,}\end{matrix}\right.
\end{align}
which, when substituted back, yields:
\begin{equation}
    \mathcal{F}_{\small\text{correlated}}\simeq\frac{8N}{\pi^2\sigma^2}\sum_{\text{odd }k}^N\frac{k^{-2}}{\frac{1-c^2}{1+c^2-2c\cos(k\pi/(N+1))}-\frac{1-c}{1+c}}.
\end{equation}
Even though for $k\approx N$ the denominator approaches $0$, the rapidly diminishing numerator $k^{-2}$ allows us to neglect this part of the sum, and assume that $\cos(k\pi/(N+1))\simeq 1$. This may seem overly hand-wavy, but such an assumption, in fact, correctly predicts the asymptotic behavior. We continue with this assumption to obtain:
\begin{align}
    \nonumber\mathcal{F}_{\small\text{correlated}}&\simeq\frac{8N}{\pi^2\sigma^2}\sum_{\text{odd }k}^N \frac{k^{-2}}{\frac{1+c}{1-c}-\frac{1-c}{1+c}}=\\\nonumber&=\frac{8N}{\pi^2\sigma^2}\frac{1-c^2}{4c}\sum_{\text{odd }k}^N k^{-2}\simeq\\&\simeq\frac{8N}{\pi^2\sigma^2}\frac{1-c^2}{4c}\frac{\pi^2}{8}=N\frac{1-c^2}{4c\sigma^2}.
\end{align}
The above argument fails for $c<0$, as the denominator approaches 0 for $k=1$, yielding CFI diverging to $\infty$.

\section{How many phases are sufficient to approximate Gaussian noise?}
\label{app:numerics}
% \subsection*{How many phases are sufficient?}

\begin{figure}[t]
    \centering
    \includegraphics[width=\linewidth]{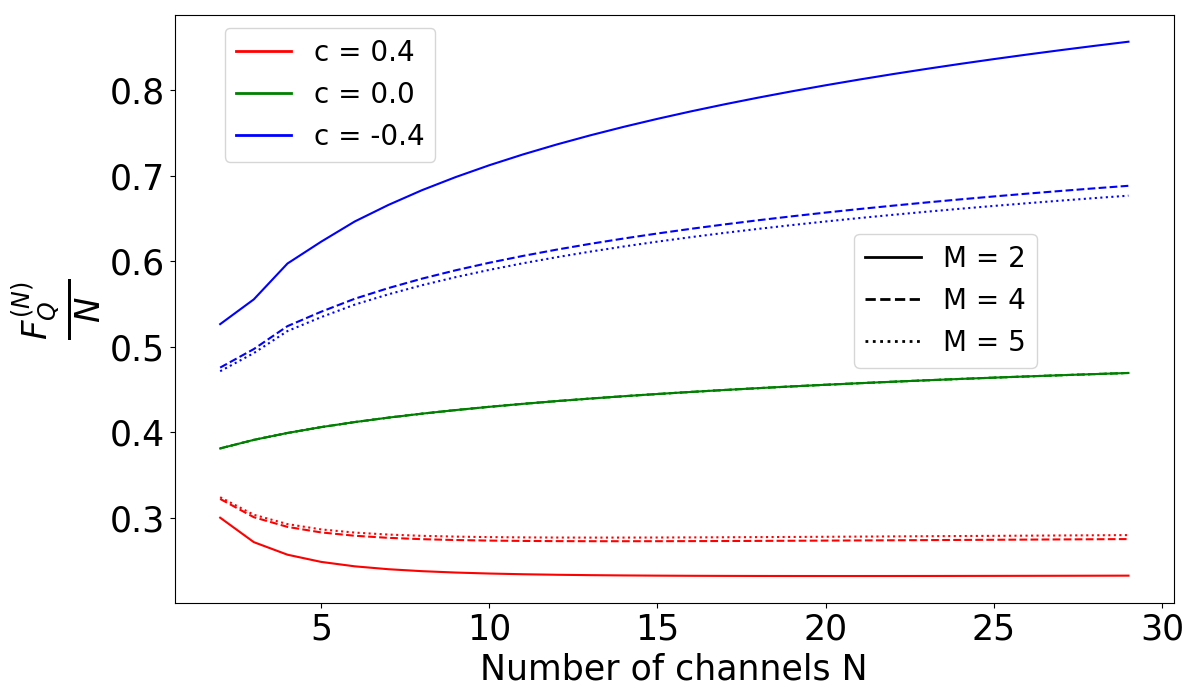}
    \caption{Fisher information per channel $F_{Q}^{(N)}/N$ with the number of channel uses $N$ with SS probe state.}

    \label{SS phase comparison}
\end{figure}

% In the main text, we discussed the optimal strategies and established the upper bound for the metrological scenario in which the quantum channels exhibit temporal correlations. To numerically analyze Markovian Gaussian noise model, we discretized it using the Rouwenhorst method, which provides a convenient and stable approximation. In our implementation, the process was approximated by $M=4$ symmetric phases.
A detailed numerical investigation of the Rouwenhorst process-driven dephasing indicates that for both moderate positive and negative correlations as well as for the uncorrelated case, the difference in the computed Fisher information between discretizations with $M = 4$ and $M = 5$ phases is negligible for all practical purposes (see Fig.~\ref{SS phase comparison}). Although increasing the value of $M$ yields a finer approximation of the continuous noise, the improvement is not substantial enough to affect the determination of the optimal strategy for the protocol. Therefore, to ensure computational efficiency while maintaining sufficient accuracy, we adopted $M = 4$ throughout the numerical simulations presented in the main text.

We should note, however, that irrespective of how many finite discrete phases one chooses, the resulting model will not be able to reconstruct properties of the continuous Gaussian model for strong enough correlations $|c| \approx 1$. In particular, for any discretized model, when correlations are strong enough, and we use enough probes, we may estimate the phase better than the phase noise discretization step, and hence be able to separate noise from the signal; see more quantitative analysis at the end of Appendix~\ref{app:higher order}.

\section{Estimation protocol with higher-order moment measurements}
\label{app:higher order}
Previously, we have discussed an estimation protocol with SS state and $J_y$ measurement, which is optimal for uncorrelated noise.
% Numerical investigations show, that for many models the SS states are still optimal probe states with correlations present \cite{Chabuda2020}, therefore we focus on optimizing the measurement.
% \textcolor{red}{(for analytical analysis of probe state optimisation we refer to appendix)}.
To generalize this protocol we introduce a \emph{multi-observable estimator} based on measured observables $\vec{X}$ with coefficients $\vec{\xi}$normalized to satisfy the local unbiasedness condition:
\begin{align}
    &\hat\theta=\frac{\vec{\xi}^T\vec{X}}{\vec{\xi}^T\partial_\theta\langle\vec{X}\rangle},\qquad\text{Var}(\hat\theta)=\frac{\vec{\xi}^T\text{VAR}(\vec{X})\vec{\xi}}{|\vec{\xi}^T\partial_\theta\langle\vec{X}\rangle|^2},\\
    &\text{VAR}(\vec{X})_{mn}=\frac{1}{2}\langle X_m X_n+X_nX_m\rangle-\langle X_m\rangle\langle X_n\rangle.
\end{align}
Minimizing the variance over coefficients $\vec{\xi}$ yields the optimal estimator for $\vec{\xi}=\text{VAR}(\vec{X})^{-1}\partial_\theta\langle\vec{X}\rangle$, with variance:
\begin{equation}
\label{eq:multivariate estimator}
    \text{Var}(\hat\theta)=(\partial_\theta\langle\vec{X}\rangle^T\text{VAR}(\vec{X})^{-1}\partial_\theta\langle\vec{X}\rangle)^{-1}.
\end{equation}
For the optimal estimator, we should include as many observables as possible in the scheme, and as $\sigma_y^{(n)}$ are the informative local measurements, we consider observables which are polynomials in variables $\sigma_y^{(n)}$ for $n\in\{1,\cdots,N\}$. For the most general translationally invariant scheme, we define observables:
\begin{equation}
    X_{t_1,\cdots,t_k}=\sum_{n=1}^{N-t_k}\sigma_y^{(n)}\sigma_y^{(n+t_1)}\cdots\sigma_y^{(n+t_k)},
\end{equation}
for $t_i$ being any strictly increasing sequence of positive integers. Although the number of such possible sequences grows with $N$, we can introduce a finite cutoff limiting the number of allowed sequences $t_k\leq t_\text{max}$, as increasing $t_\text{max}$ has diminishing influence over the scheme. By including an observable with an empty index $X_\circ=2J_y$, we see that this scheme includes the previous estimation protocol as a special case. The crucial difference is that $J_y$ is only sensitive to the proportion of positive to negative spin measurement outcomes, and therefore neglects patterns (such as clustering or repulsion of identical measurement results). For the uncorrelated case, the proportion of outcomes is a sufficient statistic, but for the correlated case we need to include patterns to obtain the optimal estimator.

% If $\langle\vec{\varphi}\rangle=\vec{0}$ then:
% \begin{align}
%     &\left\langle\prod_{k=1}^n\sigma_y^{(m_k)}\right\rangle=0\quad\text{if $n$ is odd,}\\
%      \partial_\theta&\left\langle\prod_{k=1}^n\sigma_y^{(m_k)}\right\rangle=0\quad\text{if $n$ is even.}
% \end{align}
As $\langle\vec{\varphi}\rangle=\vec{0}$, only odd moments are sensitive to parameter $\theta$, and moreover, odd moments are uncorrelated with even moments, thus making even moments unnecessary in the estimation scheme. For computing such moments, let us notice that SS states are Gaussian states, that is, spin operators can be described as components of a multivariate Gaussian distribution. 
As higher moments of Gaussian distributions can be described by first and second moments with Isserlis formula \cite{isserlis1918formula}, with moments as in Eq. (\ref{eq:OAT_a},\ref{eq:OAT_b}) we can write:
\begin{align}
    \nonumber&\left\langle\prod_{i=1}^m\sigma_{x}^{(k_i)}\prod_{j=1}^n\sigma_{y}^{(k_j)}\right\rangle_{\text{SS}}\simeq\\&\simeq\left\{\begin{matrix}
    \frac{n(n-1)}{2}\langle\sigma_x^{(k)}\rangle^m_{\tiny\text{SS}}\langle\sigma_y^{(i)}\sigma_y^{(j)}\rangle_{\tiny\text{SS}}^{n/2}&\text{if $n$ is even,}\\0&\text{if $n$ is odd,}
    \end{matrix}\right.\quad\simeq\\&\simeq\left\{\begin{matrix}
    \frac{n(n-1)}{2}\left(\frac{-1}{N}\right)^{n/2}&\text{if $n$ is even,}\\0&\text{if $n$ is odd,}
    \end{matrix}\right.
\end{align}
where all indices are pairwise different. After omitting terms of order $N^{-2}$, the only non-zero terms are for $n=0$ or $n=2$. For SS state rotated by phases $\vec{\varphi}$ we get:
\begin{align}
    % \nonumber&\left\langle\prod_{i=1}^n\sigma_y^{(k_i)}\right\rangle=\left\langle\prod_{i=1}^n\left(\sigma_y^{(k_i)}\cos\varphi_{k_i}+\sigma_x^{(k_i)}\sin\varphi_{k_i}\right)\right\rangle_{\text{OAT}}\simeq\\&\qquad\qquad\quad\simeq\prod_{i=1}^n\sin\varphi_{k_i}\left\langle\prod_{i=1}^n\sigma_x^{(k_i)}\right\rangle_{\text{OAT}}+\nonumber\\&+\sum_{i\neq j}^n\cos\varphi_{k_i}\cos\varphi_{k_j}\prod_{m\neq i,j}^n\sin\varphi_{k_m}\left\langle\sigma_y^{(k_i)}\sigma_y^{(k_j)}\prod_{m\neq i,j}^n\sigma_x^{(k_m)}\right\rangle_{\text{OAT}}\nonumber\simeq
    \nonumber&\left\langle\prod_{i=1}^n\sigma_y^{(k_i)}\right\rangle=\left\langle\prod_{i=1}^n\left(\sigma_y^{(k_i)}\cos\varphi_{k_i}+\sigma_x^{(k_i)}\sin\varphi_{k_i}\right)\right\rangle_{\text{SS}}\!\!\!\!\!\\&\simeq\prod_{i=1}^n\sin\varphi_{k_i}-\frac{1}{N}\sum_{i\neq j}^n\cos\varphi_{k_i}\cos\varphi_{k_j}\prod_{m\neq i,j}^n\sin\varphi_{k_m}.
    \label{eq:general moments}
    % \left\langle\sigma_y^{(k_i)}\sigma_y^{(k_j)}\right\rangle_{\text{OAT}}
\end{align}
% We also rewrite it to a simpler-looking form:
% \begin{equation}
%     \left\langle\prod_{i=1}^n\sigma_y^{(k_i)}\right\rangle\simeq\left(1-\frac{n+\partial_\theta^2}{2N}\right)\left(\prod_{i=1}^n\sin(\theta+\varphi_{k_i})\right)_{\theta=0},
% \end{equation}
% where $n$ factor counter-acts terms with both derivatives acting on the same function.

% For the SS probe state the general formula for qubit moments is the following:
% \begin{align}
% \label{eq:general moments}
%     % \left\langle\prod_{k=1}^n\sigma_y^{(m_k)}\right\rangle\simeq\left(1-\frac{1}{N}\partial_\theta^2\right)\left\langle\prod_{k=1}^n\sin(\theta+\varphi_{m_k})\right\rangle,
%     \left\langle\prod_{k=1}^n\sigma_y^{(m_k)}\right\rangle\simeq&\left\langle\prod_{k=1}^n\sin(\varphi_{m_k})\right\rangle+\\
%     -\frac{1}{N}\sum_{i\neq j}^n&\left\langle\cos(\varphi_{m_i})\cos(\varphi_{m_j})\prod_{k\neq i,j}^n\sin(\varphi_{m_k})\right\rangle\nonumber,
% \end{align}
% where $m_k$ are pairwise different and the second term corresponds to quantum correlations due to squeezing (we refer reader to Appendix \ref{app:SS moments} for detailed derivation).
Let us notice, that using Eq. \eqref{eq:general moments} we can express the measurement covariance matrix as a sum of two terms:
\begin{align}
    \text{VAR}(\vec{X})\simeq K(\vec{X})-\frac{1}{N}Q(\vec{X}),
\end{align}
where the leading term $K(\vec{X})$ is the covariance matrix calculated on non-squeezed, product probe state, and the correction term $Q(\vec{X})$ is due to spin-squeezing.
We define $h(m)$ as number of Pauli matrices multiplied in each term of $X_m$. Let us note, that the leading order term in $\text{VAR}(\vec{X})$ is of order $N$, so we are only interested in terms in $Q(\vec{X})$ of order $N^2$. This allows us to neglect $\sigma_y^2=1$ property in terms, where some of $\sigma_y$ operators from $X_m$ and $X_n$ act on the same qubits, leading to:
\begin{align}
    Q(\vec{X})_{mn}\simeq\frac{\tilde\partial_\theta^2}{2}\sum_{i,j}\left\langle\prod_{k=1}^{h(m)}\sin\varphi_{i+t_k^{(m)}}\prod_{l=1}^{h(n)}\sin\varphi_{j+t_l^{(n)}}\right\rangle,
    % Q(\vec{X})_{mn}=\sum_{i,j}\left\langle\prod_{k}\sin\varphi_{i+t_k^{(m)}}\prod_{l}\sin\varphi_{j+t_l^{(n)}}\right\rangle.
\end{align}
where $\tilde\partial_\theta^2$ is the second derivative with the restriction that both derivatives have to act on different functions. Now let us assume $\vec\varphi$ has exponentially vanishing correlations, so for example, processes with finite memory (Markov order). We first consider case when both derivatives act on the functions in the same product, say, without loss of generality, it is the first one, to obtain the term:
\begin{align}
    &\left|\sum_{i,j}\left\langle\frac{\tilde\partial_\theta^2}{2}\prod_{k=1}^{h(m)}\sin\varphi_{i+t_k^{(m)}}\prod_{l=1}^{h(n)}\sin\varphi_{j+t_l^{(n)}}\right\rangle\right|\leq\\&\leq \frac{h(m)^2}{2}\sum_{i,j}\left\langle\prod_{k=1}^{h(m)}\left|\sin\varphi_{i+t_k^{(m)}}\right|\prod_{l=1}^{h(n)}\left|\sin\varphi_{j+t_l^{(n)}}\right|\right\rangle\leq\nonumber\\&\leq \frac{h(m)^2}{2}\sum_{i,j}\left\langle\left|\sin\varphi_{i}\sin\varphi_{j}\right|\right\rangle\simeq\frac{h(m)^2}{2}\sum_{i,j}A|c|^{|i-j|},\nonumber
\end{align}
and due to exponential decay of correlations, the sum is of order $N$, and therefore is negligible. On the other hand, when both derivatives act on different products, we get:
\begin{align}
    &\nonumber\sum_{i,j}\left\langle\partial_\theta\left(\prod_{k=1}^{h(m)}\sin\varphi_{i+t_k^{(m)}}\right)\partial_\theta\left(\prod_{l=1}^{h(n)}\sin\varphi_{j+t_l^{(n)}}\right)\right\rangle=\\&=\sum_{i,j}\left\langle\partial_\theta\prod_{k=1}^{h(m)}\sin\varphi_{i+t_k^{(m)}}\right\rangle\left\langle\partial_\theta\prod_{l=1}^{h(n)}\sin\varphi_{j+t_l^{(n)}}\right\rangle,
\end{align}
as for $|i-j|\gg 1$ both products become uncorrelated. In conclusion, for our noise model we can take $Q(\vec{X})\simeq\partial_\theta\langle\vec{X}\rangle\partial_\theta\langle\vec{X}\rangle^T$.
% Whenever $\vec{\varphi}$ has exponentially vanishing correlations (which includes all Markov chains), in the leading order the correction matrix is a rank-1 matrix $Q(\vec{X})\simeq\partial_\theta\langle\vec{X}\rangle\partial_\theta\langle\vec{X}\rangle^T$ (derivation in Appendix \ref{app:matrix decomposition}).
We can therefore simplify Eq. \eqref{eq:multivariate estimator} by inverting the covariance matrix using the Sherman-Morrison formula \cite{sherman1950adjustment} and obtain:
% \begin{align}
%     \text{VAR}(\vec{X})&^{-1}\simeq\left(K(\vec{X})-\frac{1}{N}\partial_\theta\langle\vec{X}\rangle\partial_\theta\langle\vec{X}\rangle^T\right)^{-1}=\\
%     &=K(\vec{X})^{-1}+\frac{K(\vec{X})^{-1}\partial_\theta\langle\vec{X}\rangle\partial_\theta\langle\vec{X}\rangle^TK(\vec{X})^{-1}}{N-\partial_\theta\langle\vec{X}\rangle^T K(\vec{X})^{-1}\partial_\theta\langle\vec{X}\rangle}\nonumber,
% \end{align}
% and by substituting and simplifying:
\begin{align}
\label{eq:classical estimator}
    % (\partial_\theta\langle\vec{X}\rangle^T\text{VAR}(\vec{X})^{-1}\partial_\theta\langle\vec{X}\rangle)^{-1}
    \text{Var}(\hat\theta)\simeq (\partial_\theta\langle\vec{X}\rangle^TK(\vec{X})^{-1}\partial_\theta\langle\vec{X}\rangle)^{-1}-\frac{1}{N}.
\end{align}
We see a $-1/N$ correction to the estimator variance with SS state probe state as compared to product probe state, independently on the observables vector $\vec{X}$. These results allow for significant simplification of further computation, removing all quantum correlations from the measurement optimization problem, and allow us to directly use some results from classical estimation theory.
For example, if $\vec{\varphi}$ is multivariate Gaussian with small enough variances, we can use the approximation $\sin\varphi_k\approx \varphi_k$ during computation of moments. Under this assumption, higher-order moments don't improve the estimation scheme, as the first-order moment estimator saturates asymptotically the CR bound. Therefore, when dealing with Gaussian noise, we should expect noticeable improvement only for strong noise. Moreover, our numerical investigations show that the significant gain is obtained for $\sigma^2\gg 1$ and $c\ll 0$.
% Numerical results for such case are presented in Fig.~\ref{fig:gaussian}, and they show little benefit for including higher moments when $c>0$ and significant improvement for $c\ll0$.
% \begin{figure}[H]
%    % \centering
%     \includegraphics[width=\linewidth]{plot2.pdf}
%     \caption{Asymptotic estimator variance calculated using Eq. \eqref{eq:classical estimator} for Gaussian Markovian noise described in Eq. \eqref{eq:gaussian} for $\sigma=2$, which corresponds to $\eta=0.135$. Third order observables up to $t_\text{max}=10$ were included, as increasing cutoff and order of observables to more than 3 does not improve the estimator noticeably.}
%     \label{fig:gaussian}
% \end{figure}

\begin{figure}[t]
   % \centering
    \includegraphics[width=\linewidth]{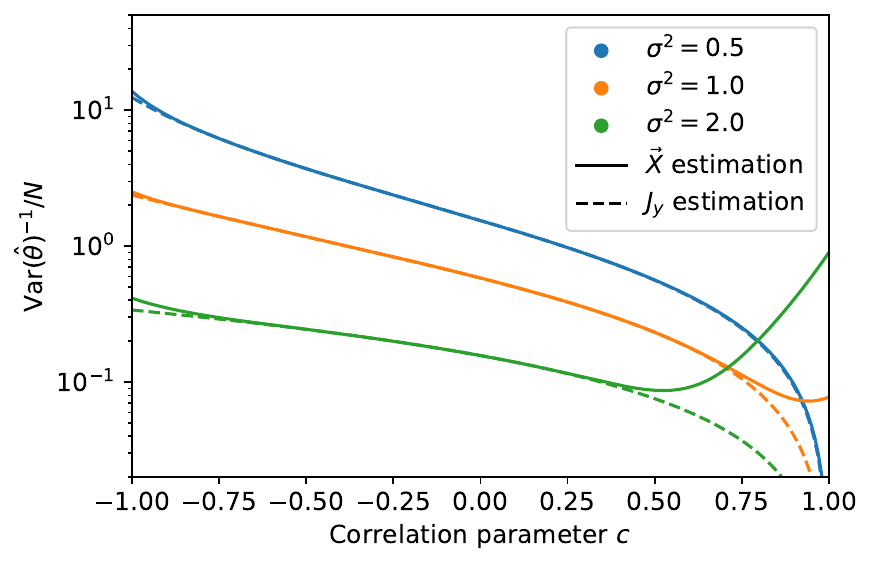}
    \caption{Asymptotic estimator variance calculated using Eq. \eqref{eq:classical estimator} for the Rouwenhorst process with $M=3$ phases state space with either $J_y$ (dashed lines) or higher moment $\vec{X}$ (solid lines) measurement. Here we use only $\vec X=(J_y,X_{1,2})^T$, as including more observables has negligible effect.}
    \label{fig:binary estimator}
\end{figure}

For the discretized Rouwenhorst process above, the argument ceases to hold, and for $c>0$ we can obtain unintuitive results:
\begin{align}
    \label{eq:limits}
    \lim_{c\to1}\lim_{N\to\infty}\frac{\mathcal{F}_Q^{(N)}}{N}\neq0,
    % \quad \lim_{\sigma\to\infty}\lim_{c\to1}\lim_{N\to\infty}\frac{\mathcal{F}_Q^{(N)}}{N}=\infty,
\end{align}
whereas for the original Markovian Gaussian process, CS bound \eqref{eq:upper} tells us, that the first limit should be equal to 0. Moreover, for $c\gg0$, increasing strength of noise decreases the estimation error, which further questions the accuracy of the approximation. 
In Fig.~\ref{fig:binary estimator} we present an example of such behaviour for $M=3$. We should stress that the discretization method ceases to be a good approximation of a continuously supported process in the strongly correlated regime. Let us note that the order of taking limits in Eq.~\eqref{eq:limits} is important, as any finite $N$ taking $c=1$ allows for Heisenberg scaling.
% and for mathematical details we refer to appendix \ref{app:rouwenhorst limit}.

\bibliographystyle{quantum}
% \begin{thebibliography}{9}
\bibliography{biblio_corr_bounds.bib}

\end{document}